\documentstyle[12pt,aasms4,epsfig]{article}
\def\in{\indent}
\def\ni{\noindent}
\def\simge{%
    \mathrel{\rlap{\raise 0.511ex
        \hbox{$>$}}{\lower 0.511ex \hbox{$\sim$}}}}
\def\simle{%
    \mathrel{\rlap{\raise 0.511ex
        \hbox{$<$}}{\lower 0.511ex \hbox{$\sim$}}}}

\begin{document}
\title{NEUTRON STAR STRUCTURE AND \\ THE EQUATION OF STATE}
\author{J. M. LATTIMER AND M. PRAKASH}

\affil{Department of Physics and Astronomy, State University of New
York at Stony Brook \\ Stony Brook, NY 11974-3800}

\begin{abstract}
The structure of neutron stars is considered from theoretical and
observational perspectives.  We demonstrate an important aspect of
neutron star structure: the neutron star radius is primarily
determined by the behavior of the pressure of matter in the vicinity
of nuclear matter equilibrium density.  In the event that extreme
softening does not occur at these densities, the radius is virtually
independent of the mass and is determined by the magnitude of the
pressure.  For equations of state with extreme softening, or those
that are self-bound, the radius is more sensitive to the mass.  Our
results show that in the absence of extreme softening, a measurement
of the radius of a neutron star more accurate than about 1 km will
usefully constrain the equation of state.  We also show that the
pressure near nuclear matter density is primarily a function of the
density dependence of the nuclear symmetry energy, while the nuclear
incompressibility and skewness parameters play secondary roles.  

In addition, we show that the moment of inertia and the binding energy
of neutron stars, for a large class of equations of state, are nearly
universal functions of the star's compactness.  These features can be
understood by considering two analytic, yet realistic, solutions of
Einstein's equations, due, respectively, to Buchdahl and Tolman.  We
deduce useful approximations for the fraction of the moment of inertia
residing in the crust, which is a function of the stellar compactness
and, in addition, the presssure at the core-crust interface. \\

\ni {\em Subject headings:} structure of stars -- equation of state --
stars: interiors -- stars: neutron 
\end{abstract}
\section{INTRODUCTION}

The theoretical study of the structure of neutron stars is crucial if
new observations of masses and radii are to lead to effective
constraints on the equation of state (EOS) of dense matter.  This
study becomes ever more important as laboratory studies may be on the
verge of yielding evidence about the composition and stiffness of
matter beyond the nuclear equilibrium density
$\rho_s\cong2.7\cdot10^{14}$ g cm$^{-3}$.  Rhoades \& Ruffini (1974)
demonstrated that the assumption of causality beyond a fiducial
density $\rho_f$ sets an upper limit to the maximum mass of a neutron
star: $4.2\sqrt{\rho_s/\rho_f}{\rm~M}_\odot$.  However, theoretical
studies of dense matter have considerable uncertainty in the
high-density behavior of the EOS largely because of the poorly
constrained many-body interactions.  These uncertainties have
prevented a firm prediction of the maximum mass of a beta-stable
neutron star.

To date, several accurate mass determinations of neutron stars are
available from radio binary pulsars (Thorsett \& Chakrabarty 1998), and
they all lie in a narrow range ($1.25-1.44$ M$_\odot$).  One neutron
star in an X-ray binary, Cyg X-2, has an estimated mass of $1.8\pm0.2$
M$_\odot$ (Orosz \& Kuulkers 1999), but this determination is not as
clean as for a radio binary.  Another X-ray binary, Vela X-1, has a reported
mass around 1.9 M$_\odot$ (van Kerkwijk et
al. 1995a), although Stickland et al. (1997) argue it to be about 1.4
M$_\odot$.  A third object, the eclipsing X-ray binary 4U 1700-37,
apparently contains an object with a mass of $1.8\pm0.4$ M$_\odot$ (Heap \&
Corcoran 1992), but Brown, Weingartner \& Wijers (1996) have argued that
since this source does not pulse and has a relatively hard X-ray
spectrum, it may contain a low-mass black hole instead.  It would
not be surprising if neutron stars in X-ray binaries had
larger masses than those in radio binaries, since the latter have
presumably accreted relatively little mass since their formation.
Alternatively, Cyg X-2 could be the first of a new and rarer
population of neutron stars formed with high masses which could
originate from more massive, rarer, supernovae.  If the high masses
for Cyg X-2 or Vela X-1 are confirmed, significant constraints on the
equation of state would be realized.

On the other hand, there is a practical, albeit theoretical, lower mass limit
for neutron stars, about $1.1-1.2$ M$_\odot$, which follows from the minimum
mass of a protoneutron star.  This is estimated by examining a
lepton-rich configuration with a low-entropy inner core of $\sim0.6$
M$_\odot$ and a high-entropy envelope (Goussard, Haensel \& Zdunik
1998).  This argument is in general agreement with the theoretical
result of supernova calculations, in which the inner homologous
collapsing core material comprises at least 1 M$_\odot$.

Although accurate masses of several neutron stars are available, a
precise measurement of the radius does not yet exist.  Lattimer
et al. (1990) (see also Glendenning 1992) have shown that the
causality constraint can be used to set a lower limit to the
radius: $R\simge3.04GM/Rc^2$.  For a 1.4 M$_\odot$ star, this is about
4.5 km.

Estimates of neutron star radii from observations have given a wide
range of results.  Perhaps the most reliable estimates stem from
observations of thermal emissions from neutron star surfaces, which
yield values of the so-called ``radiation radius''
\begin{equation}
R_\infty=R/\sqrt{1-2GM/Rc^2}\,, 
\label{rinfty}
\end{equation}
a quantity resulting from redshifting the stars luminosity and
temperature.  A given value of $R_\infty$ implies that $R<R_\infty$
and $M<0.13(R_\infty/{\rm~km})$ M$_\odot$.  Thus, a 1.4 M$_\odot$ neutron
star requires $R_\infty>10.75$ km.   Those pulsars with at least some
suspected thermal radiation generically yield effective values of
$R_\infty$ so small that it is believed that a significant part of the
radiation originates from polar hot spots rather than from the surface
as a whole.  For example, Golden \& Shearer (1999) found that upper
limits to the unpulsed emission from Geminga, coupled with a
parallactic distance of 160 pc, yielded values of $R_\infty\simle9.5$
km for a blackbody source and $R_\infty\simle10$ km for a magnetized H
atmosphere.  Similarly, Schulz (1999) estimated emission radii of less
than 5 km, assuming a blackbody for eight low mass X-ray binaries.

Other attempts to deduce a radius include analyses (Titarchuk 1994) of
X-ray bursts from sources 4U 1705-44 and 4U 1820-30 which implied
rather small values, $9.5<R_\infty<14$ km.  Recently, Rutledge et
al. (1999) found that thermal emission from neutron stars of a
canonical 10 km radius was indicated by the interburst emission.
However, the modeling of the photospheric expansion and touchdown on
the neutron star surface requires a model dependent relationship
between the color and effective temperatures.

Absorption lines in X-ray spectra have also been investigated (Inoue
1992) with a view to deducing the neutron star radius.  Candidates for
the matter producing the absorption lines are either the accreted
matter from the companion star or the products of nuclear burning in
the bursts.  In the former case, the most plausible element is thought
to be Fe, in which case the relation $R\approx3.2GM/c^2$, only
slightly larger than the minimum possible value based upon causality
(Lattimer et al. 1990; Glendenning 1992) is inferred.  In the latter
case, plausible candidates are Ti and Cr, and larger values of the
radius would be obtained.  In both cases, serious difficulties remain
in interpreting the large line widths, of order 100--500 eV, in the
$4.1 \pm 0.1$ keV line observed from many sources.

A first attempt at using light curves and pulse fractions from pulsars
to explore the $M-R$ relation suggested relatively large radii, of
order 15 km (Page 1995).  However, this method, which assumed dipolar
magnetic fields, was unable to satisfactorily reconcile the calculated
magnitudes of the pulse fractions and the shapes of the light curves
with observations.

The discovery of Quasi-Periodic Oscillations (QPOs) from X-ray
emitting neutron stars in binaries provides a possible way of limiting
neutron star masses and radii.  These oscillations are manifested as
quasi-periodic X-ray emissions, with frequencies ranging from tens to
over 1200 Hz.  Some QPOs show multiple frequencies, in particular, two
frequencies $\nu_1$ and $\nu_2$ at several hundred Hz.  These
frequencies are not constant, but tend to both increase with time
until the signal ultimately weakens and disappears.  In the beat
frequency model (Alpar \& Shaham 1985, Psaltis et al. 1998), the
highest frequency $\nu_2$ is associated with the Keplerian frequency
$\nu_K$ of inhomogeneities or blobs in an accretion disc.  The largest
such frequency measured to date is $\nu_{max}=1230$ Hz.  However,
general relativity predicts the existence of a maximum orbital
frequency, since the inner edge of an accretion disc must remain
outside of the innermost stable circular orbit, located at a radius of
$r_s=6GM/c^2$ in the absence of rotation.  This corresponds to a
Keplerian orbital frequency of $\nu_s=\sqrt{GM/r_s^3}/2\pi$ if the
star is non-rotating.  Equating $\nu_{max}$ with $\nu_s$, and since
$R<r_s$, one deduces
\begin{equation}
M\simle1.78{\rm~M}_\odot;\qquad R\simle8.86 (M/{\rm
M}_\odot){\rm~km}.\label{qpo}
\end{equation}
Corrections due to stellar rotation are straightforward to deduce and
produce small changes in these limits (Psaltis et al. 1998).  The
lower frequency $\nu_1$ is associated with a beat frequency between
$\nu_2$ and the spin frequency of the star.  This spin frequency is
large enough, of order 250-350 Hz, to alter the metric from a
Schwarzschild geometry, and increases the theoretical mass limit in
equation (\ref{qpo}) to about 2.1 M$_\odot$ (Psaltis et al. 1998).
This remains strictly an upper limit, however, unless further
observations support the interpretation that $\nu_{max}$ is associated
with orbits at precisely the innermost stable orbit.

If the frequency $\nu_2-\nu_1$ is to be associated with the spin of
the neutron star, it should remain constant in time.  However, recent
observations reveal that it changes with time in a given source.
Osherovich \& Titarchuk (1999) proposed a model in which $\nu_1$ is
the Keplerian frequency and $\nu_2$ is a hybrid frequency of the
Keplerian oscillator under the influence of a magnetospheric Coriolis
force.  In this model, the frequencies are related to the neutron star
spin frequency $\nu$ by
\begin{equation}
\nu_2=\sqrt{\nu_1^2+(\nu/2\pi)^2}.
\label{ko}
\end{equation}
Osherovich \& Titarchuk argue that this relation fits the observed
variations of $\nu_2$ and $\nu_1$ in several QPOs.  The Keplerian
frequency in Osherovich \& Titarchuk's model, being associated with
the lower frequency $\nu_1$, however, is at most 800 Hz, leading to an
upper mass limit that is nearly 3 M$_\odot$ and is therefore of little
practical use to limit either the star's mass or radius.

An alternative model, proposed by Stella \& Vietri (1999), associates
$\nu_2$ with the Keplerian frequency of the inner edge of the disc,
$\nu_K$, and $\nu_2-\nu_1$ with the precession frequency of the
periastron of slightly eccentric orbiting blobs at radius $r$ in the
accretion disc.  In a Schwarzschild geometry,
$\nu_1=\nu_K\sqrt{1-6GM/rc^2}$.  Note that
$(\nu_K-\nu_1)^{-1}$ is the timescale that an orbiting blob
recovers its original orientation relative to the neutron star and the
Earth, so that variations in flux are expected to be observed at both
frequencies $\nu_K$ and $\nu_K-\nu_1$.  Presumably, even
eccentricities of order $10^{-4}$ lead to observable effects.  This
model predicts that
\begin{equation}
\nu_1/\nu_2=1-\sqrt{1-6(GM\nu_2)^{2/3}/c^2},
\label{sv}
\end{equation}
a relation that depends only upon $M$.  Equation~(\ref{sv}) can also fit
observations of QPOs, but only if $1.9\simle M/{\rm M}_\odot\simle
2.1$.  This result is not very sensitive to complicating effects due
to stellar rotation: the Lense-Thirring effect and oblateness.  This
mechanism only depends on gravitometric effects, and may apply also to
accreting black hole systems (Stella, Vietri \& Morsink 1999).

Prospects for a radius determination have improved in recent years
with the discovery of a class of isolated, non-pulsing, neutron stars.
The first of these is the nearby object RX J185635-3754, initially
discovered in X-rays (Walter, Wolk \& Neuha\"user 1996) and confirmed
with the Hubble Space Telescope (Walter \& Matthews 1997). The
observed X-rays, from the ROSAT satellite, are consistent with
blackbody emission with an effective temperature of about 57 eV and
relatively little extinction.  The fortuitous location of the star, in
the foreground of the R CrA molecular cloud, coupled with the small
levels of extinction, limits the distance to $D<120$ pc.  The fact
that the source is not observable in radio and its lack of variability
in X-rays implies that it is not a pulsar, unlike other identified
radio-silent isolated neutron stars.  This gives the hope that the
observed radiation is not contaminated with non-thermal emission as in
the case for pulsars.  

The X-ray flux of RX~J185635-3754, combined with a best-fit blackbody
$T_{eff}=57$ eV, yields $R_\infty\approx 7.3 (D/120 {\rm~pc}){\rm~
km}$.  Such a value for $R_\infty$, even coupled with the maximum
distance of 120 pc, is too small to be consistent with any neutron
star with more than 1 M$_\odot$.  But the optical flux is about a
factor of 4 brighter than what is predicted by the X-ray blackbody.
The reconciliation the X-ray and optical fluxes through atmosphere
modeling naively implies an increase in $R_\infty$ of approximately
$4^{2/3}\cong2.5$.  (This results since the X-ray flux is proportional
to $R_\infty^2T_{eff}^4$, while the optical flux is on the
Rayleigh-Jeans tail of the spectrum and is hence proportional to
$R_\infty^2T_{eff}$.  One seeks to enhance the predicted optical flux
by 4 while keeping the X-ray flux fixed, as this is approximately
equal to the total flux.) An et al. (2000) determined for
non-magnetized heavy element atmospheres that $R_\infty/D\cong
0.18\pm0.05$ km pc$^{-1}$, which is rough agreement with the above
naive expectations.  However, uncertainties in the atmospheric
composition and the quality of the existing data precluded obtaining a
more precise estimate of $R_\infty/D$.  An et al. concluded, in
agreement with expectations based upon the general results of Romani
(1987) and Rajagopal, Romani \& Miller (1997), 
that the predicted spectrum of a
heavy element atmosphere, but not a light element atmosphere, was
consistent with all the observations.  This is in contrast to the
conclusions of Pavlov et al. (1996), whose results for RX J185635-3754
implied that the observations in the optical and X-ray bands were
incompatible with atmospheric modelling for both heavy element and
light element non-magnetized atmospheres, unless the distance to this
star is greater than the presumed maximum of 120 pc based upon the
star's location in front of the R Cor Aus molecular cloud.  Future
prospects for determining the radius of this neutron star are
discussed in \S 7.

Our objectives in this paper are 1) to demonstrate specifically how
the accurate measurement of a neutron star radius would constrain the
dense matter EOS, and 2) to provide general relationships for other
structural quantities, such as the moment of inertia and the binding
energy, that are relatively EOS-independent, and which could be used
to constrain the neutron star mass and/or radius.  We will examine a
wide class of equations of state, including those that have extreme
softening at high densities.  In addition, we will examine analytic
solutions to Einstein's equations which shed light on the results we
deduce empirically.  In all cases, we will focus on non-rotating,
non-magnetized neutron stars at zero temperature.

Lindblom (1992) had suggested that a series of mass and radius
measurements would be necessary to accurately constrain the dense
matter equation of state.  His technique utilizes a numerical
inversion of the neutron star structure equation.  Our results instead
suggest that important constraints on the EOS can be achieved with
even a single radius measurement, if it is accurate enough, and that
the quality of the constraint is not very sensitive to the mass.  The
fact that the range of accurately determined neutron star masses is so
small, only about 0.2 M$_\odot$ to date, further implies that
important constraints can be deduced without simulaneous mass-radius
measurements.  Of course, several measurements of neutron star masses
and radii would greatly enhance the constraint on the equation of
state.

In \S~2, the equations of state selected in this paper are discussed.
In \S~3, the mass-radius relation for a sample of these equations of
state are discussed.  A quantitative relationship between the radii of
normal neutron stars and the pressure of matter in the vicinity of
$n_s$ is empirically established and theoretically justified.  In
turn, how the matter's pressure at these densities depends upon
fundamental nuclear parameters is developed.  In \S~4, analytic
solutions to the general relativistic equations of hydrostatic
equilibrium are explored.  These lead to useful approximations for
neutron star structure and which directly correlate other observables
such as moments of inertia and binding energy to the mass and radius.
It is believed that the distribution of the moment of inertia inside
the star is crucial in the interpretation of glitches observed in the
spin down of pulsars, so that measurements of the sizes and
frequencies of glitches can constrain neutron star masses and radii
(Link, Epstein \& Lattimer 1999).  In \S~5, expressions for the
fraction of moment of inertia contained within the stellar crust, as a
function of mass, radius, and equation of state, are derived.  In
\S~6, expressions for the binding energy are derived.  \S~7 contains a
summary and outlook.

\section{EQUATIONS OF STATE}

\def\lsim{\mathrel{\rlap{
\lower3pt\hbox{\hskip-3pt$\sim$}}
\raise1pt\hbox{$<$}}}

The composition of a neutron star chiefly depends on the nature of
strong interactions, which are not well understood in dense matter.
Most models that have been investigated can be conveniently grouped
into three broad categories: nonrelativistic potential models,
relativistic field theoretical models, and relativistic
Dirac-Brueckner-Hartree-Fock models.  In each of these approaches, the
presence of additional softening components such as hyperons, Bose
condensates or quark matter, can be incorporated.  Details of these
approaches have been further considered in Lattimer et al. (1990) and
Prakash et al. (1997).  A representative sample, and some general
attributes, including references and typical compositions, of
equations of state employed in this paper are summarized in Table
1.

We have used four equations of state taken from Akmal \& Pandharipande
(1998).  These are: AP1 (the AV18 potential), AP2 (the AV18 potential
plus $\delta v_b$ relativistic boost corrections), AP3 (the AV18
potential plus a three-body UIX potential ), and AP4 (the AV18
potential plus the UIX potential plus the $\delta v_b$ boost).  Three
equations of state from M\"uller \& Serot (1996), labelled MS1--3,
correspond to different choices of the parameters $\xi$ and $\zeta$
which determine the strength of the nonlinear vector and isovector
interactions at high densities.  The numerical values used are
$\xi=\zeta=0; \xi=1.5, \zeta=0.06$; and $\xi=1.5, \zeta=0.02$,
respectively.  Six EOSs from the phenomenological non-relativistic
potential model of Prakash, Ainsworth \& Lattimer (1988), labelled
PAL1--6, were chosen, which have different choices of the symmetry
energy parameter at the saturation density, its density dependence,
and the bulk nuclear matter incompressibility parameter $K_s$.  The
incompressibilities of PAL1--5 were chosen to be $K_s=180$ or 240 MeV,
but PAL6 has $K_s=120$ MeV.  Three interactions from the
field-theoretical model of Glendenning \& Moszkowski (1991) are taken
from their Table II; in order, they are denoted GM1--3.  Two
interactions from the field-theoretical model of Glendenning \&
Schaffner-Bielich (1999) correspond, in their notation, to GL78 with
$U_K(\rho_0)=-140$ MeV and TM1 with $U_K=-185$ MeV.  The labels
denoting the other EOSs in Table I are identical to those in the
original references.

The rationale for exploring a wide variety of EOSs, even some that are
relatively outdated or in which systematic improvements are performed,
is two-fold.  First, it provides contrasts among widely different
theoretical paradigms.  Second, it illuminates general relationships
that exist between the pressure-density relation and the macroscopic
properties of the star such as the radius.  For example, AP4
represents the most complete study to date of Akmal \& Pandharipande
(1998), in which many-body and special relativistic corrections are
progressively incorporated into prior models, AP1--3.  AP1--3 are
included here because they represent different pressure-energy
density-baryon density relations, and serve to reinforce correlations
between neutron star structure and microscopic physics observed using
alternative theoretical paradigms.  Similarly, several different
parameter sets for other EOSs are chosen.

In all cases, except for PS (Pandharipande \& Smith 1975), the
pressure is evaluated assuming zero temperature and beta equilibrium
without trapped neutrinos.  PS only contains neutrons among the
baryons, there being no charged components.  We chose to include this
EOS, despite the fact that it has been superceded by more
sophisticated calculations by Pandharipande and coworkers, because it
represents an extreme case producing large radii neutron stars.

The pressure-density relations for some of the selected EOSs are shown
in Figure \ref{fig:P-rho}.  There are two general classes of equations
of state.  First, {\em normal} equations of state have a pressure
which vanishes as the density tends to zero.  Second, {\em self-bound}
equations of state have a pressure which vanishes at a significant
finite density.

The best-known example of self-bound stars results from Witten's
(1984) conjecture (also see Fahri \& Jaffe 1984, Haensel, Zdunik \&
Schaeffer 1986, Alcock \& Olinto 1988, and Prakash et al. 1990) that
strange quark matter is the ultimate ground state of matter.  In this
paper, the self-bound EOSs are represented by strange-quark matter
models SQM1--3, using perturbative QCD and an MIT-type bag model, with
parameter values given in Table 2.  The existence of an energy ceiling
equal to the baryon mass, 939 MeV, for zero pressure matter requires
that the bag constant $B\le94.92$ MeV fm$^{-3}$.  This limiting value
is chosen, together with zero strange quark mass and no interactions
($\alpha_c=0$), for the model SQM1.  The other two models chosen, SQM2
and SQM3, have bag constants adjusted so that their energy ceilings
are also 939 MeV.

For normal matter, the EOS is that of an interacting nucleon gas above
a transition density of 1/3 to 1/2 $n_s$.  Below this density, the
ground state of matter consists of heavy nuclei in equilibrium with a
neutron-rich, low-density gas of nucleons.  In general, a
self-consistent evaluation of the equilibrium that exists below the
transition density, and the evaluation of the transition density
itself, has been carried out for only a few equations of state (e.g.,
Bethe, Pethick \& Sutherland 1972, Negele \& Vautherin 1974, Lattimer
et al. 1985; Lattimer and Swesty 1990).  We have therefore not
plotted the pressure below about 0.1 MeV fm$^{-3}$ in Figure
\ref{fig:P-rho}.  For densities $0.001 < n < 0.08$ fm$^{-3}$ we employ
the EOS of Negele \& Vautherin (1974), while for densities $n<0.001$
fm$^{-3}$ we employ the EOS of Bethe, Pethick \& Sutherland (1972).
However, for most of the purposes of this paper, the pressure in the
region $n<0.1$ fm$^{-3}$ is not relevant as it does not significantly
affect the mass-radius relation or other global aspects of the star's
structure.  Nevertheless, the value of the transition density, and the
pressure there, are important ingredients for the determination of the
size of the superfluid crust of a neutron star that is believed to be
involved in the phenomenon of pulsar glitches (Link, Epstein \&
Lattimer 1999).

There are three significant features to note in Figure \ref{fig:P-rho}
for normal EOSs.  First, there is a fairly wide range of predicted
pressures for beta-stable matter in the density domain $n_s/2<n<2n_s$.
For the EOSs displayed, the range of pressures covers about a factor
of five, but this survey is by no means exhaustive.  That such a wide
range in pressures is found is somewhat surprising, given that each of
the EOSs provides acceptable fits to experimentally-determined nuclear
matter properties.  Clearly, the extrapolation of the pressure from
symmetric matter to nearly pure neutron matter is poorly constrained.
Second, the {\em slopes} of the pressure curves are rather similar.  A
polytropic index of $n\simeq1$, where $P=Kn^{1+1/n}$, is implied.
Third, in the density domain below $2n_s$, the pressure-density
relations seem to fall into two groups.  The higer pressure group is
primarily composed of relativistic field-theoretical models, while the
lower pressure group is primarily composed of non-relativistic potential
models.  As we show in \S~3, the pressure in the vicinity of $n_s$ is
mostly determined by the symmetry energy properties of the EOS, and it
is significant that relativistic field-theoretical models generally
have symmetry energies that increase proportionately to the density
while potential models have much less steeply rising symmetry
energies.

A few of the plotted normal EOSs have considerable softening at high
densities, especially PAL6, GS1, GS2, GM3, PS and PCL2.  PAL6 has an
abnormally small value of incompressibility ($K_s=120$ MeV).  GS1 and
GS2 have phase transitions to matter containing a kaon condensate, GM3
has a large population of hyperons appearing at high density, PS
has a phase transition to a neutral pion condensate and a neutron
solid, and PCL2 has a phase transition to a mixed phase containing
strange quark matter.  These EOSs can be regarded as representative of
the many suggestions of the kinds of softening that could occur at
high densities.

\section{NEUTRON STAR RADII}

Figure \ref{fig:M-R} displays the mass-radius relation for cold,
catalyzed matter using these EOSs.  The causality constraint described
earlier and contours of $R_\infty$ are also indicated in Figure
\ref{fig:M-R}.  With the exception of model GS1, the EOSs used to
generate Figure~\ref{fig:M-R} result in maximum masses greater than 1.442
M$_\odot$, the limit obtained from PSR 1913+16.  From a theoretical
perspective, it appears that values of $R_\infty$ in the range of
12--20 km are possible for normal neutron stars whose masses are
greater than 1 M$_\odot$.

Corresponding to the two general types of EOSs, there are two general
classes of neutron stars.  {\em Normal} neutron stars are
configurations with zero density at the stellar surface and which have
minimum masses, of about 0.1 M$_\odot$, that are primarily determined
by the EOS below $n_s$.  At the minimum mass, the radii are generally
in excess of 100 km.  The second class of stars are the so-called
{\em self-bound} stars, which have finite density, but zero pressure, at
their surfaces.  They are represented in Figure \ref{fig:M-R} by 
strange quark matter stars (SQM1--3).

Self-bound stars have no minimum mass, unlike the case of normal
neutron stars for which pure neutron matter is unbound.  Unlike normal
neutron stars, the maximum mass self-bound stars have nearly the
largest radii possible for a given EOS.  If the strange quark mass
$m_s=0$ and interactions are neglected ($\alpha_c=0$), the maximum
mass is related to the bag constant $B$ in the MIT-type bag model by
$M_{max}=2.033~(56{\rm~MeV~fm}^{-3}/B)^{1/2}~{\rm M}_\odot$.  Prakash
et al. (1990) and Lattimer et al. (1990) showed that the addition of a
finite strange quark mass and/or interactions produces larger maximum
masses.  The constraint that $M_{max}>1.44$ M$_\odot$ is thus
automatically satisfied for all cases by the condition that the energy
ceiling is 939 MeV.  In addition, models satisfying the energy ceiling
constraint, with any values of $m_s$ and $\alpha_c$, have larger radii
for every mass than the case SQM1.  For the MIT model, the locus of
maximum masses of self-bound stars is given simply by $R\cong1.85
R_s$~(Lattimer et al. 1990), where $R_s=2GM/Rc^2$ is the Schwarzschild
radius, which is shown in the right-hand panel of
Figure~\ref{fig:M-R}.  Strange quark stars with electrostatically
supported normal-matter crusts~(Glendenning \& Weber 1992) have larger
radii than those with bare surfaces.  Coupled with the additional
constraint $M>1{\rm M}_\odot$ from protoneutron star models, MIT-model strange
quark stars cannot have $R<8.5$ km or
$R_\infty<10.5$ km.  These values are comparable to the possible
lower limits for a Bose (pion or kaon) condensate EOS.

Although the $M-R$ trajectories for normal stars can be strikingly
different, in the mass range from 1 to 1.5 M$_\odot$ or more it is
usually the case that the radius has relatively little dependence upon
the stellar mass.  The major exceptions illustrated are the model GS1,
in which a mixed phase containing a kaon condensate appears at a
relatively low density and the model PAL6 which has an extremely small
nuclear incompressibility (120 MeV).  Both of these have considerable
softening and a large increase in central density for $M>1$ M$_\odot$.
Pronounced softening, while not as dramatic, also occurs in models GS2
and PCL2, which contain mixed phases containing a kaon condensate and
strange quark matter, respectively.  All other normal EOSs in this
figure, except PS, contain only baryons among the hadrons.

While it is generally assumed that a stiff EOS implies both a large
maximum mass and a large radius, many counter examples exist.  For
example, GM3, MS1 and PS have relatively small maximum masses but have
large radii compared to most other EOSs with larger maximum masses.
Also, not all EOSs with extreme softening have small radii for $M>1$
M$_\odot$ (e.g., GS2, PS).  Nonetheless, for stars with masses greater than
1 M$_\odot$, only models with a large degree of softening (including
strange quark matter configurations) can have
$R_\infty<12$ km.  Should the radius of a neutron star ever be
accurately determined to satisfy $R_\infty<12$ km, a strong case could
be made for the existence of extreme softening.

To understand the relative insensitivity of the radius to the mass for
normal neutron stars, it is relevant that a Newtonian polytrope with
$n=1$ has the property that the stellar radius is independent of both
the mass and central density.  Recall that most EOSs, in the density
range of $n_s-2n_s$, have an effective polytropic index of about one (see
Figure \ref{fig:P-rho}).  An $n=1$ polytrope also has the property that the
radius is proportional to the square root of the constant $K$ in the
polytropic pressure law $P=K\rho^{1+1/n}$.  This suggests that there
might be a quantitative relation between the radius and the pressure
that does not depend upon the EOS at the highest densities, which
determines the overall softness or stiffness (and hence, the maximum
mass).

In fact, this conjecture may be verified.  Figure~\ref{fig:P-R} shows
the remarkable empirical correlation which exists between the radii of
1 and 1.4 M$_\odot$ normal stars and the matter's pressure evaluated
at fiducial densities of 1, 1.5 and 2 $n_s$.  Table~1 explains the EOS
symbols used in Figure~\ref{fig:P-R}.  Despite the relative
insensitivity of radius to mass for a particular EOS in this mass
range, the nominal radius $R_M$, which is defined as the radius at a
particular mass $M$ in solar units, still varies widely with the EOS
employed.  Up to $\sim 5$ km differences are seen in $R_{1.4}$, for
example.  Of the EOSs in Table 1, the only severe violations of this
correlation occurs for PCL2 and PAL6 at 1.4 M$_\odot$ for $n_s$, and
for PS at both 1 and 1.4 M$_\odot$ for $2n_s$.  In the case of PCL2,
this is relatively close to the maximum mass, and the matter has
extreme softening due to the existence of a mixed phase with quark
matter.  (A GS model intermediate between GS1 and GS2, with a maximum
mass of 1.44 M$_\odot$, would give similar results.)  In the case of
PS, it is clear from Figure~\ref{fig:P-rho} that extensive softening
occurs already by $1.5n_s$.  We emphasize that this correlation is
valid only for cold, catalyzed neutron stars, i.e., not for
protoneutron stars which have finite entropies and might contain
trapped neutrinos.

Numerically, the correlation has the form of a power law:
\begin{equation}
R_M \simeq C(n,M)~[P(n)]^{0.23-0.26}\,,
\label{correl}
\end{equation}
where $P(n)$ is the total pressure inclusive of leptonic contributions
evaluated at the density $n$, and $C(n,M)$ is a number that depends on
the density $n$ at which the pressure was evaluated and the stellar
mass $M$.  An exponent of 1/4 was chosen for display in
Figure~\ref{fig:P-R}, but the correlation holds for a small range of
exponents about this value.  Using an exponent of 1/4, and ignoring
points associated with EOSs with phase transitions in the density
ranges of interest, we find values for $C(n,M)$, in units of km
fm$^{3/4}$ MeV$^{-1/4}$, which are listed in Table 3.  The error bars
are taken from the standard deviations.
The correlation is seen to be somewhat tighter for the baryon density
$n=1.5 n_s$ and $2 n_s$ cases.

The fact that the exponent is considerably less than the Newtonian
value of 1/2 can be quantitatively understood by considering a relativistic
generalization of the $n=1$ polytrope due to Buchdahl (1967).  He
found that the EOS
\begin{equation}
\rho=12\sqrt{p_*P}-5P\,,\label{buch}
\end{equation}
where $p_*$ is a constant fiducial pressure independent of density,
has an analytic solution of Einstein's equations.  This solution is
characterized by the quantities $p_*$ and $\beta\equiv GM/Rc^2$, and
the stellar radius is found to be
\begin{equation}
R=(1-\beta)c^2\sqrt{\pi\over288p_*G(1-2\beta)}\,.
\label{p*}
\end{equation}
For completeness, we summarize below the metric functions, the
pressure and the mass-energy density as functions of coordinate radius
$r$:
\begin{eqnarray}
e^\nu &\equiv& g_{tt}=
(1-2\beta)(1-\beta-u)(1-\beta+u)^{-1}\,;\cr 
e^\lambda &\equiv& g_{rr}=
(1-2\beta)(1-\beta+u)(1-\beta-u)^{-1}(1-\beta+\beta\cos Ar^\prime)^{-2}\,;\cr
8\pi PG/c^4 &=& A^2u^2(1-2\beta)(1-\beta+u)^{-2}\,;\cr
8\pi\rho G/c^2&=& 2A^2u(1-2\beta)(1-\beta-3u/2)(1-\beta+u)^{-2}\,. 
\label{buch1}
\end{eqnarray}
where
\begin{eqnarray}
r &=& r^\prime(1-\beta+u)(1-2\beta)^{-1}\,;\cr
u &=& \beta(Ar^\prime)^{-1}\sin Ar^\prime\,;\cr
A^2 &=& 288\pi p_*Gc^{-4}(1-2\beta)^{-1}.
\label{buch2}
\end{eqnarray}
Note that $R\propto p_*^{-1/2}(1+\beta^2/2+\dots)$, so for a given
value of $p_*$, the radius increases very slowly with mass.

To estimate the exponent, it is instructive to analyze the response of
$R$ to a change of pressure at some fiducial density $\rho$, for a
fixed mass $M$.  (At the relatively low densities of interest, the
difference between using $n$ or $\rho$ in the following analysis is
not significant.)  We find the exponent to be
\begin{eqnarray}
{d\ln R\over d\ln P}\Biggr|_{\rho,M} &=& {d\ln R\over d\ln
p_*}\Biggr|_\beta {d\ln p_*\over d\ln P}\Biggr|_{\rho}\Biggl[1+{d\ln
R\over d\ln\beta}\Biggr|_{p_*}\Biggr]^{-1} \cr
 &=& \frac 12 \Biggl(1-{5\over6}\sqrt{P\over p_*}\Biggr)
{(1-\beta)(1-2\beta)\over(1-3\beta+3\beta^2)}\,. 
\end{eqnarray}
In the limit $\beta\rightarrow0$, one has $P\rightarrow0$ and the
exponent $d\ln R/d\ln P\Bigr|{\rho, M}\rightarrow1/2$, the value
characteristic of an $n=1$ Newtonian polytrope. Finite values of
$\beta$ and $P$ render the exponent smaller than 1/2.  If the
stellar mass and radius are about 1.4 M$_\odot$ and 15 km,
respectively, for example, equation~(\ref{p*}) gives
$p_*=\pi/(288 R^2)\approx4.85\cdot10^{-5}$ km$^{-2}$ (in geometrized
units).  Furthermore, if the fiducial density is $\rho\approx
1.5m_bn_s\approx2.02\cdot10^{-4}$ km$^{-2}$ (also in geometrized
units, with $m_b$ the baryon mass), equation~(\ref{buch}) implies that in
geometrized units $P\approx8.5\cdot10^{-6}$ km$^{-2}$.  Since the
value of $\beta$ in this case is 0.14, one then obtains $d\ln R/d\ln
P\simeq0.31$.  This result, while mildly sensitive to the choices for
$\rho$ and $R$, provides a reasonable explanation of the correlation,
equation~(\ref{correl}).  The fact that the exponent is smaller than 1/2 is
clearly an effect due to general relativity.

The existence of this correlation is significant because the pressure
of degenerate neutron-star matter near the nuclear saturation density
$n_s$ is, in large part, determined by the symmetry properties of the
EOS, as we now discuss.  Thus, the measurement of a neutron star
radius, if not so small as to indicate extreme softening, could
provide an important clue to the symmetry properties of matter.  In
either case, valuable information will be obtained.

Studies of pure neutron matter strongly suggest that the specific
energy of nuclear matter near the saturation density may be expressed
as an expansion quadratic in the asymmetry $(1-2x)$, where $x$ is the
proton fraction, which can be terminated after only one term (Prakash,
Ainsworth \& Lattimer 1988).  In this case, the energy per particle
and pressure of cold, beta stable nucleonic matter is
\begin{eqnarray}
E(n,x) &\simeq& E(n,1/2) + S_v(n)(1-2x)^2   \,, \nonumber \\
P(n,x) &\simeq& n^2[E^\prime(n,1/2)+ S_v^\prime (1-2x)^2] \,,
\label{enuc}
\end{eqnarray}
where $E(n,1/2)$ is the energy per particle of symmetric matter and
$S_v(n)$ is the bulk symmetry energy (which is density dependent).
Primes denote derivatives with respect to density. At $n_s$, the
symmetry energy can be estimated from nuclear mass systematics and has
the value $S_v\equiv S_v(n_s) \approx 27-36~{\rm MeV}$.  Attempts
to further restrict this range from consideration of fission
barriers and the energies of giant resonances have led to ambiguous
results.  Both the magnitude of $S_v$ and its density dependence
$S_v(n)$ are currently uncertain.  Part of the symmetry energy is due
to the kinetic energy for noninteracting matter, which for degenerate
nucleonic matter is proportional to $n^{2/3}$, but the remainder of
the symmetry energy, due to interactions, is also expected to
contribute significantly to the overall density dependence.

Leptonic contributions must to be added to equation~(\ref{enuc}) to obtain the
total energy and pressure; the electron energy per baryon is $(3/4)\hbar
cx(3\pi^2nx)^{1/3}$.
Matter in neutron stars is in beta equilibrium, i.e., $\mu_e =
\mu_n - \mu_p = - \partial E/\partial x$, which permits the evaluation
of the equilibrium proton fraction and the total pressure 
$P$ may be written at a particular density
in terms of fundamental nuclear parameters (Prakash 1996).  
For example, the pressure at the saturation density is simply
\begin{eqnarray}
P_s=n_s(1-2x_s)[n_sS_v^\prime(1-2x_s)+S_v x_s]\,,
\end{eqnarray}
where $S_v^\prime\equiv(\partial S_v(n)/\partial n)_{n=n_s}$
and the equilibrium proton fraction at $n_s$ is
\begin{eqnarray}
x_s\simeq(3\pi^2 n_s)^{-1}(4S_v/\hbar c)^3 \simeq 0.04\,,
\end{eqnarray}
for $S_v=30$ MeV. Due to the small value of $x_s$, we find that
$P_s\simeq n_s S_v^\prime$.  The inclusion of muons, which generally
begin to appear around $n_s$, does not qualitatively affect these results.

Were we to evaluate the pressure at a larger density, contributions
featuring other nuclear
parameters, including the nuclear incompressibility $K_s=9(dP/dn)|n_s$
and the skewness $K_s^\prime=-27n_s^3(d^3E/dn^3)|_{n_s}$, also become
significant.  For analytical purposes, the nuclear matter energy per
baryon, in MeV, may be expanded in the vicinity of $n_s$ as
\begin{eqnarray}
E(n,1/2) = -16 + \frac {K_s}{18} \left(\frac {n}{n_s}-1 \right)^2 
- \frac {K_s^\prime}{27}  \left(\frac {n}{n_s}-1 \right)^3\,.
\label{eexp}
\end{eqnarray}
Experimental constraints to the compression modulus $K_s$, most
importantly from analyses of giant monopole resonances (Blaizot et
al. 1995; Youngblood et al. 1999), give $K_s\cong 220$ MeV.  The
skewness parameter $K_s^\prime$ has been estimated to lie in the range
1780--2380 MeV (Pearson 1991, Rudaz et al. 1992), but in these calculations
contributions from the surface symmetry energy were neglected.
For values of $K_s^\prime$ this large, equation~(\ref{eexp}) cannot be used
beyond about 1.5$n_s$.  Evaluating the pressure for
$n=1.5n_s$, we find
\begin{eqnarray}
P(1.5n_s)= 2.25n_s [K_s/18-K_s^\prime/216 + n_s(1-2x)^2S_v^\prime] \,.
\end{eqnarray}
Assuming that $S_v(n)$ is approximately proportional to the density,
as it is in most relativistic field theoretical models,
$S_v^\prime(n)\cong S_v/n_s$.  Since the $K_s$ and $K_s^\prime$ terms
largely cancel, the symmetry term comprises most of the total.
Once again, the result that the pressure is mostly sensitive to the
density dependence of the symmetry energy is found.

The sensitivity of the radius to the symmetry energy can further
demonstrated by the parametrized EOS of PAL (Prakash, Ainsworth \&
Lattimer 1988).  The symmetry energy function $S_v(n)$ is a direct
input in this parametrization and can be chosen to reproduce the
results of more microscopic calculations.  Figure~\ref{fig:alp9} shows
the dependence of mass-radius trajectories as the quantities $S_v$ and
$S_v(n)$ are alternately varied.  Clearly, of the two variations, the
density dependence of $S_v(n)$ is more important in determining the
neutron star radius.  Note also the weak sensitivity of the maximum
neutron star mass to $S_v$, and that the maximum mass depends more
strongly upon the function $S_v(n)$.

At present, experimental guidance concerning the density dependence of
the symmetry energy is limited and mostly based upon the division of
the nuclear symmetry energy between volume and surface contributions.
Upcoming experiments involving heavy-ion collisions which might sample
densities up to $\sim (3-4)n_s$, will be limited to analyzing
properties of the nearly symmetric nuclear matter EOS through a study
of matter, momentum, and energy flow of nucleons.  Thus, studies of
heavy nuclei far off the neutron drip lines using radioactive ion
beams will be necessary in order to pin down the properties of the
neutron-rich regimes encountered in neutron stars.

\section{MOMENTS OF INERTIA}

Besides the stellar radius, other global attributes of neutron stars
are potentially observable, including the moment of inertia and the
binding energy.  These quantities depend primarily upon the ratio
$M/R$ as opposed to details of the EOS, as can be readily seen by
evaluating them using analytic solutions to Einstein's equations.
Although over 100 analytic solutions to Einstein's equations are known
(Delgaty \& Lake 1998), nearly all of them are physically unrealistic.
However, three analytic solutions are of particular interest in
normal neutron star structure.

The first is the well-known Schwarzschild interior solution for an
incompressible fluid, $\rho=\rho_c$, where $\rho$ is the mass-energy
density.  This case, hereafter referred to as ``Inc'', is mostly of
interest because it determines the minimum compactness $\beta=GM/Rc^2$
for a neutron star, namely 4/9, based upon the central pressure being
finite.  Two aspects of the incompressible fluid that are physically
unrealistic, however, include the fact that the sound speed is
everywhere infinite, and that the density does not vanish on the
star's surface.

The second analytic solution, due to Buchdahl (1967), is described in
equation~(\ref{buch1}).  We will refer to this solution as ``Buch''.

The third analytic solution (which we will refer to as ``T VII'') was
discovered by Tolman (1939) and corresponds to the case when the
mass-energy density $\rho$ varies quadratically, that is,
\begin{equation}
\rho=\rho_c[1-(r/R)^2].
\end{equation}
Of course, this behavior is to be expected at both extremes
$r\rightarrow0$ and $r\rightarrow R$.  However, this is also an
eminently reasonable representation for intermediate regions, as
displayed in Figure~\ref{fig:prof}, which contains results for neutron
stars more massive than 1.2 M$_\odot$.  A wide variety of EOSs are
sampled in this figure, and they are listed in Table~1.

Because the T VII solution is often overlooked in the literature (for
exceptions, see, for example, Durgapal \& Pande 1980 and Delgaty \&
Lake 1998), it is summarized here.  It is useful in establishing
interesting and simple relations that are insensitive to the EOS.  In
terms of the variable $x=r^2/R^2$ and the compactness parameter
$\beta=GM/Rc^2$, the assumption $\rho=\rho_c(1-x)$ results in
$\rho_c=15\beta c^2/(8\pi GR^2)$.  The solution of Einstein's
equations for this density distribution is:
\begin{eqnarray}
e^{-\lambda} &=& 1-\beta x(5-3x)\,,\qquad e^\nu =
(1-5\beta/3)\cos^2\phi\,, \cr P &=& {c^4\over4\pi R^2
G}\left[\sqrt{3\beta
e^{-\lambda}}\tan\phi-{\beta\over2}(5-3x)\right]\,, \qquad n= {(\rho
c^2+P)\over m_bc^2}{\cos\phi\over\cos\phi_1}\,, \cr \phi &=&
(w_1-w)/2+\phi_1\,,\qquad w =
\log\left[x-5/6+\sqrt{e^{-\lambda}/(3\beta)}\right]\,,\cr \phi_c &=&
\phi(x=0)\,, \quad \phi_1 =\phi(x=1)=
\tan^{-1}\sqrt{\beta/[3(1-2\beta)]}\,, \quad w_1 = w(x=1)\,.
\end{eqnarray}
The central values of $P/\rho c^2$ and the square of the sound speed
$c_s^2$ are
\begin{equation}
{P\over\rho c^2}\Biggr|_c={2\over15}\sqrt{3\over\beta}\Bigr({c_{s}\over
c}\Bigr)^2\,,\quad \Bigr({c_{s}\over 
c}\Bigr)^2=\tan\phi_c\left(\tan\phi_c+\sqrt{\beta\over3}\right)\,.
\end{equation}
This solution, like that of Buchdahl's, is scale-free, with the
parameters $\beta$ and $\rho_c$ (or $M$ and $R$).  There are obvious
limitations to the range of parameters for realistic models: when
$\phi_c=\pi/2$, or $\beta\approx0.3862$, $P_c$ becomes infinite, and
when $\beta\approx0.2698$, $c_{s}$ becomes causal ({i.e., $c$).
Recall that for an incompressible fluid, $P_c$ becomes infinite when
$\beta=4/9$, and this EOS is acausal for all values of $\beta$.  For
the Buchdahl solution, $P_c$ becomes infinite when $\beta=2/5$ and the
causal limit is reached when $\beta=1/6$.  For comparison, the causal
limit for realistic EOSs is $\beta\cong0.33$ (Lattimer et al. 1990,
Glendenning 1992), as previously discussed.

The general applicability of these exact solutions can be gauged by analyzing
the moment of inertia, which, for a star uniformly
rotating with angular velocity $\Omega$, is
\begin{equation}
I=(8\pi/3)\int_0^R r^4(\rho+P/c^2)e^{(\lambda-\nu)/2}
(\omega/\Omega) dr\,.
\label{inertia}
\end{equation}
The metric function $\omega$ is a solution of the equation
\begin{equation}
d[r^4e^{-(\lambda+\nu)/2}\omega^\prime]/dr + 4r^3\omega
de^{-(\lambda+\nu)/2}/dr=0
\label{diffomeg}
\end{equation}
with the surface boundary condition
\begin{equation}\omega_R=\Omega-{R\over3}\omega^\prime_R
=\Omega\left[1-{2GI\over R^3c^2}\right].
\label{boundary}
\end{equation}
The second equality in the above follows from the definition of $I$ and the TOV
equation.  Writing $j=\exp[-(\nu+\lambda)/2]$, the
TOV equation becomes
\begin{equation}
j^\prime=-4\pi Gr(P/c^2+\rho)je^\lambda/c^2\,.
\end{equation}
Then, one has
\begin{equation}
I=-{2c^2\over3G}\int {\omega\over\Omega}r^3dj =
{c^2R^4\omega^\prime_R\over6G\Omega} \,. \end{equation}

Unfortunately, an
analytic representation of $\omega$ or the moment of inertia for any of the
three exact solutions is not available.  However, approximations which are
valid in the causal regime to within 0.5\% are
\begin{eqnarray}
I_{Inc}/MR^2 &\simeq&  2(1-0.87\beta-0.3\beta^2)^{-1}/5\,,\label{iinc} \\
I_{Buch}/MR^2 &\simeq& 
(2/3-4/\pi^2)(1-1.81\beta+0.47\beta^2)^{-1}\,,\label{ibuc} \\
I_{T VII}/MR^2 &\simeq& 2(1-1.1\beta-0.6\beta^2)^{-1}/7\,.\label{itol}
\end{eqnarray}
In each case, the small $\beta$ limit gives the corresponding
Newtonian result.  Figure~\ref{mominert} indicates that the T VII
approximation is a rather good approximation to most EOSs without
extreme softening at high densities, for $M/R\ge0.1$ M$_\odot$/km.
The EOSs with softening fall below this trajectory.  
Ravenhall \& Pethick (1994) had suggested the expression
\begin{equation}
I_{RP}/MR^2\simeq0.21/(1-2\beta)
\end{equation}
as an approximation for the moment of inertia; however, we find
that this expression is not a good overall fit, as shown in
Figure~\ref{mominert}.  

For low-mass stars, none of the analytic approximations are suitable,
and the moment of inertia deviates substantially from the behavior of
an incompressible fluid.  Although neutron stars of such small mass
are unlikely to exist, it is interesting to examine the behavior of
$I$ in the limit of small compactness, especially the suprising result
that $I/MR^2\rightarrow0$ as $\beta\rightarrow0$.  It is well known
from the work of Baym, Bethe \& Pethick (1971) that the adiabatic
index of matter below nuclear density is near to, but less than 4/3.
As the compactness parameter $\beta$ decreases, a greater fraction of
the star's mass lies below $n_s$.  To the extent that these stars can
be approximated as polytropes (i.e., having a constant polytropic
index $n$), Table 4 shows how the quantity $I/MR^2$ varies with $n$.
For a polytropic index of 3, corresponding to an adiabatic exponent of
4/3, $I/MR^2\simeq0.075$, considerably lower than the value of 0.4 for
an incompressible fluid.  Calculations of matter at subnuclear density
agree on the fact that the adiabatic exponent of matter further
decreases with decreasing density, until the neutron drip point (near
$4.3\times10^{11}$ g cm$^{-3}$) is approached and the exponent is near
zero.  Although the central densities of minimum mass neutron stars
are about $2\times10^{14}$ g cm$^{-3}$, much of the mass of the star
is at considerably lower density, unlike the situation for solar
mass-sized neutron stars which are relatively centrally condensed.
Thus, as $\beta$ decreases, the quantity $I/MR^2$ rapidly decreases,
approaching the limiting value of zero as an effective polytropic
index of nearly 5 is achieved.

Another interesting result from Figure~\ref{mominert} concerns the
moments of inertia of strange quark matter stars.  Such stars are
relatively closely approximated by incompressible fluids, this
behavior becoming exact in the limit of $\beta\rightarrow0$.  This
could have been anticipated from the $M\propto R^3$ behavior of the
$M-R$ trajectories for small $\beta$ strange quark matter stars as
observed in Figure~\ref{fig:M-R}.

\section{CRUSTAL FRACTION OF THE MOMENT OF INERTIA}

A new observational constraint involving $I$ concerns pulsar
glitches.  Occasionally, the spin rate of a pulsar will suddenly
increase (by about a part in $10^6$) without warning after years of
almost perfectly predictable behavior.  However, Link, Epstein \&
Lattimer (1999) argue that these glitches are not completely random:
the Vela pulsar experiences a sudden spinup about every three years,
before returning to its normal rate of slowing.  Also, the size of a
glitch seems correlated with the interval since the previous glitch,
indicating that they represent self-regulating instabilities for which
the star prepares over a waiting time.  The angular momentum
requirements of glitches in Vela imply that $\ge 1.4$\% of the
star's moment of inertia drives these events.

Glitches are thought to represent angular momentum transfer between
the crust and another component of the star. In this picture, as a
neutron star's crust spins down under magnetic
torque, differential rotation develops between the stellar crust and
this component. The more rapidly rotating component then acts as an
angular momentum reservoir which occasionally exerts a spin-up torque
on the crust as a consequence of an instability.  A popular notion at
present is that the freely spinning component is a superfluid flowing
through a rigid matrix in the thin crust, the region in which
dripped neutrons coexist with nuclei, of the star.  As the solid
portion is slowed by electromagnetic forces, the liquid continues to
rotate at a constant speed, just as superfluid He continues to spin
long after its container has stopped.  This superfluid is usually
assumed to locate in the star's crust, which thus must contain at least
1.4\% of the star's moment of inertia.  

The high-density boundary of the crust is naturally set by the phase
boundary between nuclei and uniform matter, where the pressure is
$P_t$ and the density $n_t$.  The low-density boundary is the neutron
drip density, or for all practical purposes, simply the star's surface
since the amount of mass between the neutron drip point and the
surface is negligible.  One can utilize equation~(\ref{inertia}) to
determine the moment of inertia of the crust alone with the
assumptions that $P/c^2<<\rho$, $m(r)\simeq M$, and $\omega
j\simeq\omega_R$ in the crust.  Defining $\Delta R$ to be the crust
thickness, that is, the distance between the surface and the point
where $P=P_t$,
\begin{equation}
\Delta I\simeq{8\pi\over3}{\omega_R\over\Omega}\int_{R-\Delta
R}^R \rho r^4e^\lambda dr\simeq
{8\pi\over3GM}{\omega_R\over\Omega}\int_0^{P_t}r^6dP\,,
\label{deltaip}
\end{equation}
where $M$ is the star's total mass and the TOV equation was used in
the last step.  In the crust, the
fact that the EOS is of the approximate polytropic form $P\simeq
K\rho^{4/3}$ can be used to find an approximation for the integral
$\int r^6dP$, {\em viz.}
\begin{equation}
\int_0^{P_t}r^6dP\simeq P_tR^6\left[1+
{2P_t\over n_t m_nc^2}{(1+7\beta)(1-2\beta)\over\beta^2}\right]^{-1}\,.
\end{equation}
For most neutron stars, the approximation equation~(\ref{itol}) gives
$I$ in terms of $M$ and $R$, and equation~(\ref{boundary}) gives
$\omega_R/\Omega$ in terms of $I$ and $R$, the quantity $\Delta I/I$
can thus be cast as a function of $M$ and $R$ with the only
dependences upon the EOS arising from the values of $P_t$ and $n_t$;
there is no explicit dependence upon the EOS at any other density.
However, the major dependence is mostly upon the value of $P_t$, since
$n_t$ enters only as a correction.  We then find
\begin{equation}{\Delta I\over I}\simeq{28\pi P_t
R^3\over3 
Mc^2}{(1-1.67\beta-0.6\beta^2)\over\beta}\left[1+{2P_t(1+5\beta-14\beta^2)\over 
n_t
m_bc^2\beta^2}\right]^{-1}.
\label{dii}
\end{equation} 

In general, the EOS parameter $P_t$, in the units of MeV fm$^{-3}$,
varies over the range $0.25<P_t<0.65$ for realistic EOSs.  The
determination of this parameter requires a calculation of the
structure of matter containing nuclei just below nuclear matter
density that is consistent with the assumed nuclear matter EOS.
Unfortunately, few such calculations have been performed.  Like the
fiducial pressure at and above nuclear density which appears in
equation~(\ref{correl}), $P_t$ should depend sensitively upon the behavior
of the symmetry energy near nuclear density.

Since the calculation of the pressure below nuclear density has not
been consistently done for most realistic EOSs, we arbitrarily choose
$n_t=0.07$ fm$^{-3}$ and compare the approximation equation~(\ref{dii})
with the results of full structural calculations in
Figure~\ref{fig:M-R-2}.  Two extreme values of $P_t$ were assumed in
the full structural calculations to identify the core-crust boundary.
Irrespective of this choice, the agreement between the analytical
estimate equation~(\ref{dii}) and the full calculations appears to be
good for all EOSs, including ones with extreme softening.  We
also note that Ravenhall \& Pethick (1994) developed a different, but
nearly equivalent, analytic formula for the quantity $\Delta I/I$ as a
function of $M, R, P_t$ and $\mu_t$, where $\mu_t$ is the neutron
chemical potential at the core-crust phase boundary.  This prediction
is also displayed in Figure~\ref{fig:M-R-2}.

Link, Epstein \& Lattimer (1999) established a lower limit to the
radius of the Vela pulsar by using equation~(\ref{dii}) with $P_t$ at
its maximum value and the glitch constraint $\Delta I/I\ge0.014$.  A
minimum radius can be found by combining this constraint with the
largest realistic value of $P_t$ from any equation of state, namely
about 0.65 MeV fm$^{-3}$.  Stellar models that are compatible with
this constraint must fall to the right of the $P_t=0.65$ MeV fm$^{-3}$
contour in Figure~\ref{fig:M-R-2}.  This imposes a constraint upon the
radius, which is approximately equivalent to
\begin{equation}
R>3.9+3.5 M/{\rm M}_\odot-0.08 (M/{\rm M}_\odot)^2{\rm~km}\,. 
\label{glitch}
\end{equation}
As shown in the figure,
this constraint is somewhat more stringent than one based upon causality.
Better estimates of the maximum value of $P_t$ should make this
constraint more stringent.

\section{BINDING ENERGIES}

The binding energy formally represents the energy gained by assembling
$N$ baryons.  If the baryon mass is $m_b$, the binding energy is
simply $BE=Nm_b-M$ in mass units.  However, the quantity $m_b$ has
various interpretations in the literature.  Some authors take it to be
939 MeV/$c^2$, the same as the neutron or proton mass.  Others take it
to be about 930 MeV/$c^2$, corresponding to the mass of C$^{12}$/12 or
Fe$^{56}$/56.  The latter choice would be more appropriate if $BE$ was
to represent the energy released in by the collapse of a
white-dwarf-like iron core in a supernova explosion.  The difference
in these definitions, 10 MeV per baryon, corresponds to a shift of
$10/939\simeq0.01$ in the value of $BE/M$.  This energy, $BE$, can be deduced 
from neutrinos detected from a supernova
event; indeed, it might be the most precisely determined aspect of the
neutrino signal.

Lattimer \& Yahil (1989) suggested that the binding energy could be
approximated as
\begin{equation}
BE\approx 1.5\cdot10^{51} (M/{\rm M}_\odot)^2 {\rm~ergs} = 0.084
(M/{\rm M}_\odot)^2 {\rm~M}_\odot\,.
\label{lybind}
\end{equation}
Prakash et al. (1997) also concluded that such a formula was a
reasonable approximation, based upon a comparison of selected
non-relativistic potential and field-theoretical 
models.  In Figure~\ref{bind}, this formula is compared to exact
results, which shows that it is accurate at best to about $\pm20$\%.
The largest deviations are for stars with extreme softening or large mass.

Here, we propose a more accurate representation of the binding
energy:
\begin{equation}
BE/M \simeq 0.6\beta/(1-0.5\beta)\,, \label{newbind}
\end{equation}
which incorporates some radius dependence.  Thus, the observation of supernova
neutrinos, and the estimate of the total radiated neutrino energy, will yield
more accurate information about $M/R$ than about $M$ alone.

In the cases of the incompressible fluid and the Buchdahl solution, analytic
results for the binding energy can be found:
\begin{eqnarray}
BE_{Inc}/M &=& {3\over4\beta}\Bigl({\sin^{-1}\sqrt{2\beta}\over
\sqrt{2\beta}}-\sqrt{1-2\beta}\Bigr)-1\approx{3\beta\over5}+{9\beta^2\over14}+{5
\beta^3\over6}+\cdots\,; \\
BE_{Buch}/M &=&  
(1-1.5\beta)(1-2\beta)^{-1/2}(1-\beta)^{-1}-1\approx{\beta\over2}+{\beta^2\over2
}+{3\beta^3\over4}+\cdots\,.
\end{eqnarray}
In addition, an expansion for the T VII solution can be found:
\begin{eqnarray}
BE_{T VII}/M \approx
{11\beta\over21}+{7187\beta^2\over18018}+{68371\beta^3\over306306}+\cdots\,.
\end{eqnarray}
The exact results for the three analytic solutions of Einstein's
equations, as well as the fit of equation~(\ref{newbind}), are
compared to some EOSs in Figure~\ref{bind1}.  It can be seen
that for stars without extreme softening both the T VII and Buch
solutions are rather realistic.  However, for EOSs with softening, the
deviations from this can be substantial.  Thus, until information
about the existence of softening in neutron stars is available, the
binding energy alone provides only limited information about the
star's structure or mass.

\section{SUMMARY AND OUTLOOK}

We have demonstrated the existence of a strong correlation between the
pressure near nuclear saturation density inside a neutron star and the
radius which is relatively insensitive to the neutron star's mass and
equation of state for normal neutron stars.  In turn, the pressure
near the saturation density is primarily determined by the isospin
properties of the nucleon-nucleon interaction, specifically, as
reflected in the density dependence of the symmetry energy, $S_v(n)$.
This result is not sensitive to the other nuclear parameters such as
$K_s$, the nuclear incompressibility parameter, or $K^\prime_s$, the
skewness parameter.  This is important, because the value of the
symmetry energy at nuclear saturation density and the density
dependence of the symmetry energy are both difficult to determine in
the laboratory.  Thus, a measurement of a neutron star's radius would
yield important information about these quantities.

Any measurement of a radius will have some intrinsic uncertainty.  In
addition, the empirical relation we have determined between the
pressure and radius has a small uncertainty.  It is useful to display
how accurately the equation of state might be established from an
eventual radius measurement.  This can be done by inverting equation
(\ref{correl}), which yields
\begin{equation}
P(n) \simeq [R_M/C(n,M)]^4\,.
\label{correli}
\end{equation}
The inferred ranges of pressures, as a function of density and for
three possible values of $R_{1.4}$, are shown in Figure~\ref{fig:err}.
It is assumed that the mass is 1.4 M$_\odot$, but the results are
relatively insensitive to the actual mass.  Note from Table 3 that the
differences between $C$ for 1 and 1.4 M$_\odot$ are typically less than the
errors in $C$ itself.  The light shaded areas show the pressures
including only errors associated with $C$.  The dark shaded areas show
the pressures when a hypothetical observational error of 0.5 km is
also taken into account.  These results suggest that a useful
restriction to the equation of state is possible if the radius of a
neutron star can be measured to an accuracy better than about 1 km.

The reason useful constraints might be obtained from just a single
measurement of a neutron star radius, rather than requiring a series
of simultaneous mass-radii measurements as Lindblom (1992) proposed,
stems from the fact that we have been able to establish the empirical
correlation, equation (\ref{correl}).  In turn, it appears that this
correlation exists because most equations of state have slopes $d \ln
P/d \ln n\simeq2$ near $n_s$.

The best prospect for measuring a neutron star's radius may be
the nearby object RX J185635-3754.  It is anticipated that parallax
information for this object will be soon available (Walter, private
communication).  In addition, it may be possible to identify spectral
lines with the Chandra and XMM X-ray facilities that would not only
yield the gravitational redshift, but would identify the atmospheric
composition.  Not only would this additional information reduce the
uncertainty in value of $R_\infty$, but, {\em both} the mass and
radius for this object might thereby be determined.  It is also
possible that an estimate of the surface gravity of the star can be
found from further comparisons of observations with atmospheric
modelling, and this would provide a further check on the mass and
radius.

We have presented simple expressions for the moment of inertia, the
binding energy, and the crustal fraction of the moment of inertia for
normal neutron stars which are largely independent of the EOS.  
If the magnitudes of observed glitches from Vela are connected with the 
crustal fraction of moment of inertia, the formula we derived establishes 
a more stringent limit on the radius than causality.

We thank A. Akmal, V. R. Pandharipande and J. Schaffner-Bielich for
making the results of their equation of state calculations available
to us in tabular form.  This work was supported in part by the USDOE
grants DOE/DE-FG02-87ER-40317 \& DOE/DE-FG02-88ER-40388.
\section*{REFERENCES}

\ni Akmal, A. \& Pandharipande, V.R. 1997, Phys. Rev., C56, 2261 \\
\ni Alcock, C. \& Olinto, A. 1988, Ann. Rev. Nucl. Sci., 38, 161 \\
\ni Alpar, A. \& Shaham, J. 1985, Nature, 316, 239 \\
\ni An, P., Lattimer, J.M., Prakash, M. \& Walter, F. 2000, in preparation \\
\ni Baym, G., Pethick, C. J. \& Sutherland, P. 1971, ApJ, 170, 299 \\
\ni Blaizot, J.P., Berger, J.F.,  Decharg\'{e}, J. \&
Girod, M. 1995, Nucl. Phys., A591, 431 \\
\ni Brown, G. E., Weingartner, J. C. \& Wijers, R. A. M. J. 1996, ApJ, 463, 
297 \\
\ni Buchdahl, H.A. 1967, ApJ, 147, 310 \\
\ni Delgaty, M.S.R.  \& Lake, K. 1998, Computer Physics Communications,
115, 395 \\
\ni Durgapal, M.C.  \& Pande, A.K.  1980, J. Pure \& Applied Phys., 18, 171 \\
\ni Engvik, L., Hjorth-Jensen, M., Osnes, E., Bao, G. \&
\O stgaard, E. 1994, \\ \in Phys. Rev. Lett., 73, 2650 \\
\ni ------1996, ApJ, 469, 794 \\
\ni Fahri, E. \& Jaffe, R. 1984, Phys. Rev., D30, 2379 \\
\ni Friedman, B. \& Pandharipande, V.R. 1981, Nucl. Phys., A361, 502 \\
\ni  Glendenning, N.K. 1992, Phys. Rev., D46, 1274 \\
\ni Glendenning, N.K.  \& Moszkowski, S.A. 1991, Phys. Rev. Lett.,
67, 2414 \\
\ni Glendenning, N.K.  \& Schaffner-Bielich, J. 1999, Phys. Rev., C60, 
025803 \\ 
\ni Glendenning, N.K. \& Weber, F. 1992, ApJ, 400, 672 \\
\ni Golden, A. \& Shearer, A. 1999, A\&A, 342, L5 \\
\ni Goussard J.-O., Haensel, P. \& Zdunik, J. L. 1998, A\&A, 330, 1005. \\
\ni Haensel, P., Zdunik, J. L. \& Schaeffer, R. 1986, A\&A, 217, 137 \\
\ni Heap, S. R. \& Corcoran, M. F. 1992, ApJ, 387, 340 \\
\ni Inoue, H. 1992, in {The Structure and Evolution of Neutron Stars},
ed. D. Pines, \\ \in R. Tamagaki and S. Tsuruta, (Redwood City:
Addison Wesley, 1992) 63 \\
\ni Lattimer, J.M.,  Pethick, C.J.,  Ravenhall, D.G. \& Lamb, D.Q.
1985, Nucl. Phys., A432, \\ \in 646 \\
\ni Lattimer, J.M., Prakash, M., Masak, D. \& Yahil, A. 1990,
ApJ, 355, 241 \\
\ni Lattimer, J.M. \& Swesty, F.D. 1991, Nucl. Phys., A535, 331 \\
\ni Lattimer, J.M.  \&  Yahil, A. 1989, ApJ, 340, 426 \\
\ni Link, B., Epstein, R.I. \& Lattimer, J.M. 1999, Phys. Rev. Lett.,
83, 3362\\
\ni Lindblom, L. 1992, ApJ, 398, 569 \\
\ni M\"uller, H.  \& Serot, B.D. 1996, Nucl. Phys., A606, 508 \\
\ni M\"uther, H., Prakash, M. \& Ainsworth, T.L. 1987, Phys. Lett., 
B199, 469 \\
\ni Negele, J. W. \& Vautherin, D. 1974, Nucl. Phys., A207, 298 \\
\ni Orosz, J. A. \& Kuulkers, E. 1999, MNRAS, 305, 132 \\
\ni Osherovich, V. \& Titarchuk, L. 1999, ApJ, 522,
L113 \\
\ni Page, D. 1995, ApJ, 442, 273 \\
\ni Pavlov, G.G., Zavlin, V.E., Truemper, J.
\& Neuhauser, R. 1996, ApJ, 472, L33 \\
\ni Pandharipande, V. R. \& Smith, R. A. 1975, Nucl. Phys., A237, 507 \\
\ni Pearson, J.M. 1991, Phys. Lett., B271, 12 \\
\ni Prakash, M. 1996, in {Nuclear Equation of State\/},
ed. A. Ansari \& L. Satpathy (Singapore: \\ \in World Scientific), p. 229 \\
\ni Prakash, M., Ainsworth, T.L. \& Lattimer, J.M. 1988,
Phys. Rev. Lett., 61, 2518 \\
\ni Prakash, Manju., Baron, E.  \& Prakash, M. 1990, Phys. Lett., B243, 175 \\
\ni Prakash, M., Bombaci, I, Prakash, Manju., Lattimer, J.M.,
Ellis, P.J. \& Knorren, R. 1997, \\ \in Phys. Rep., 280, 1 \\
\ni Prakash, M., Cooke, J.R. \& Lattimer, J.M. 1995, Phys. Rev., D52, 661 \\
\ni Psaltis, D. et al. 1998, ApJ, 501, L95 \\
\ni Rajagopal, M., Romani, R.W. \& Miller, M. C.  1997, ApJ, 479, 347  \\
\ni Ravenhall, D.G. \& Pethick, C.J. 1994, ApJ, 424, 846 \\
\ni Rhoades, C. E. \& Ruffini, R. 1974, Phys. Rev. Lett., 32, 324 \\
\ni Romani, R.W. 1987, ApJ, 313, 718 \\
\ni Rudaz, S., Ellis, P. J., Heide, E. K. and Prakash, M. 1992, 
Phys. Lett., B285, 183 \\
\ni Rutledge, R., Bildstein, L., Brown, E., Pavlov, G. G. \& Zavlin,
V. E. AAS Meeting \#193, \\ \in Abstract \#112.03 \\
\ni Schulz, N. S. 1999, ApJ, 511, 304 \\
\ni Stella, L. \& Vietri, M. 1999, Phys. Rev. Lett., 82, 17  \\
\ni Stella, L., Vietri, M. \& Morsink, S. 1999, ApJ, 524, L63 \\
\ni Stickland, D. Lloyd, C. \& Radzuin-Woodham, A. 1997, MNRAS, 286, L21 \\
\ni Thorsett, S.E. \& Chakrabarty, D. 1999, ApJ, 512, 288 \\
\ni Titarchuk, L. 1994,  ApJ, 429, 340 \\
\ni Tolman, R.C. 1939, Phys. Rev., 55, 364 \\
\ni van Kerkwijk, J. H., van Paradijs, J. \& Zuiderwijk,
E. J. 1995, A\&A, 303, 497 \\
\ni Walter, F., Wolk, S. \& Neuha\"user, R. 1996, Nature, 379, 233  \\
\ni Walter, F. \& Matthews, L. D. 1997, Nature, 389, 358  \\
\ni Wiringa, R.B., Fiks, V. \& Fabrocine, A. 1988, Phys. Rev., C38. 1010 \\
\ni Witten, E. 1984, Phys. Rev., D30, 272 \\
\ni Youngblood, D.H.,  Clark, H.L. \& Lui, Y.-W. 1999, Phys. Rev. Lett.,
82, 691 \\

\newpage

\begin{center}
\centerline{TABLE 1}
\vspace*{0.15in}
\centerline{EQUATIONS OF STATE  }
\begin{tabular}{l|l|l|l} \hline\hline
Symbol & Reference & Approach & Composition \\ \hline
FP  & Friedman \& Pandharipande (1981) & Variational & np \\
PS & Pandharipande \& Smith (1975) & Potential & n$\pi^0$ \\
WFF(1-3) & Wiringa, Fiks \& Fabrocine (1988) & Variational & np \\
AP(1-4) & Akmal \& Pandharipande (1998) & Variational & np \\
MS(1-3) & M\"uller \& Serot (1996) & Field Theoretical & np \\   
MPA(1-2) & Mu\"ther, Prakash \& Ainsworth (1987) & Dirac-Brueckner HF & np \\
ENG & Engvik et al. (1996) & Dirac-Brueckner HF & np \\
PAL(1-6)  & Prakash, Ainsworth \& Lattimer (1988) & Schematic Potential & np \\
GM(1-3) & Glendenning \& Moszkowski (1991) & Field Theoretical & npH \\
GS(1-2) & Glendenning \& Schaffner-Bielich (1999) & Field Theoretical & npK\\
PCL(1-2) & Prakash, Cooke \& Lattimer (1995) & Field Theoretical & npHQ
\\ 
SQM(1-3) & Prakash, Cooke \& Lattimer (1995) & Quark Matter & Q $(u,d,s)$\\
\hline
\end{tabular}
\end{center}
\vspace*{0.15in}
NOTES.---- Approach refers to the
underlying theoretical technique.  Composition refers to strongly
interacting components (n=neutron, p=proton, H=hyperon, K=kaon,
Q=quark); all models include leptonic contributions.

\newpage
\begin{center}
\centerline{TABLE 2}
\vspace*{0.15in}
\centerline{PARAMETERS FOR SELF-BOUND STRANGE QUARK STARS}
\begin{tabular}{l|c|c|c} \hline\hline
Model & $B$ (MeV fm$^{-3})$ & $m_s$ (MeV) & $\alpha_c$ \\ \hline
SQM1 & 94.92 & 0 & 0 \\
SQM2 & 64.21 & 150 & 0.3 \\
SQM3 & 57.39 & 50 & 0.6 \\ \hline
\end{tabular}
\end{center}
\vspace*{0.15in} NOTES.---- Numerical values employed in the MIT bag
model as described in Fahri \& Jaffe (1984).

\vspace*{1.5in}
\begin{center}
\centerline{TABLE 3}
\vspace*{0.15in}
\centerline{THE QUANTITY $C(n,M)$ OF EQUATION \ref{correl}}
\begin{tabular}{l|l|l}\hline\hline
$n$ & 1 M$_\odot$ & 1.4 M$_\odot$ \\ \hline
$n_s$ & $9.53\pm0.32$ & $9.30\pm0.60$ \\
$1.5n_s$ & $7.14\pm0.15$ & $7.00\pm0.31$ \\
$2n_s$ & $5.82\pm0.21$ & $5.72\pm0.25$ \\ \hline
\end{tabular}
\end{center}
\vspace*{0.15in} NOTES.---- The quantity $C(n,M)$, in units of km
fm$^{3/4}$ MeV$^{-1/4}$, which relates the pressure (evaluated at density
$n$) to the radius of neutron stars of mass $M$.  The errors are
standard deviations.}

\newpage
\begin{center}
\centerline{TABLE 4}
\vspace*{0.15in}
\centerline{MOMENTS OF INERTIA FOR POLYTROPES }
\begin{tabular}{l|l|l|l}\hline\hline
Index $n$ & $I/MR^2$ & Index $n$ & $I/MR^2$ \\ \hline
0   & 0.4  & 3.5 & 0.045548 \\
0.5 & 0.32593 & 4.0 & 0.022573 \\
1.0 & 0.26138 & 4.5 & 0.0068949 \\ 
1.5 & 0.20460 & 4.8 & 0.0014536 \\
2.0 & 0.15485  & 4.85 & 0.00089178  \\
2.5 & 0.11180  & 4.9 & 0.0004536 \\
3.0 & 0.075356  & 5.0 & 0 \\ \hline
\end{tabular}
\end{center}
\vspace*{0.15in}
NOTES.---- The quantity $I/MR^2$ for polytropes, which satisfy 
the relation $P=K\rho^{1+1/n}$ ($\rho$ is the mass-energy density), 
as a function of the polytropic index $n$. 
\newpage
\section*{FIGURE CAPTIONS}

\ni FIG. 1.----The pressure-density relation for a selected
set of EOSs contained in Table 1.  The pressure is in units of
MeV fm$^{-3}$ and the density is in units of baryons per cubic fermi.
The nuclear saturation density is approximately $0.16$ fm$^{-3}$.

\vspace*{.2in}

\ni FIG. 2.----Mass-radius curves for several EOSs listed in Table 1.
The left panel is for stars containing nucleons and, in some cases,
hyperons.  The right panel is for stars containing more exotic
components, such as mixed phases with kaon condensates or strange
quark matter, or pure strange quark matter stars.  In both panels, the
lower limit causality places on $R$ is shown as a dashed line, a
constraint derived from glitches in the Vela pulsar is shown as the
solid line labelled $\Delta I/I=0.014$, and contours of constant
$R_\infty=R/\sqrt{1-2GM/Rc^2}$, are shown as dotted
curves. In the right panel, the theoretical trajectory of maximum
masses and radii for pure strange quark matter stars is marked by the
dot-dash curve labelled $R=1.85R_s$.

\vspace*{0.2in}

\ni FIG. 3.----~Empirical relation between pressure, in units of MeV fm$^{-3}$,
and radius, in km, for EOSs listed in Table 1.  The upper panel shows
results for 1 M$_\odot$ (gravitational mass) stars; the lower panel is
for 1.4 M$_\odot$ stars.  The different symbols show values of
$RP^{-1/4}$ evaluated at three fiducial densities.

\vspace*{0.2in}

\ni FIG. 4.----~Left panel: Mass-radius curves for selected PAL (Prakash,
Ainsworth \& Lattimer 1988) forces showing the sensitivity to symmetry
energy.  The left panel shows variations arising from different
choices of $S_v$, the symmetry energy evaluated at $n_s$; the right
panel shows variations arising from different choices of $S_v(n)$, the
density dependent symmetry energy.  In this figure, the shorthand
$u=n/n_s$ is used.

\vspace*{0.2in}

\ni FIG. 5.----~Profiles of mass-energy density ($\rho$), relative to central
values ($\rho_c$), in neutron stars for several EOSs listed in Table 1.
For reference, the thick black lines show the simple quadratic
approximation $1-(r/R)^2$.

\vspace*{0.2in}

\ni FIG. 6.----~The moment of inertia $I$, in units of $MR^2$, for several
EOSs listed in Table 1.  The curves labelled ``Inc'', ``T VII'', ``Buch''
and ``RP'' 
are for an incompressible fluid, the Tolman (1939) VII solution,  the
Buchdahl (1967) solution,
and an approximation of Ravenhall \& Pethick (1994), 
respectively.  The inset shows details of $I/MR^2$ for $M/R \rightarrow 0$.

\vspace*{0.2in}

\ni FIG. 7.----~Mass-radius curves for selected EOSs from Table 1, comparing
theoretical contours of $\Delta I/I=0.014$ from approximations
developed in this paper, labelled ``LP'', and from Ravenhall \&
Pethick (1994), labelled ``RP'', to numerical results (solid dots).
Two values of $P_t$, the transition pressure demarking the
crust's inner boundary, which bracket estimates in the literature, are
employed.  The region to the left of the $P_t=0.65$ MeV fm$^{-3}$
curve is forbidden if Vela glitches are due to angular momentum transfers
between the crust and core, as discussed in Link, Epstein \& Lattimer
(1999).  For comparison, the region excluded by causality alone lies
to the left of the dashed curve labelled ``causality'' as determined
by Lattimer et al. (1990) and Glendenning (1992).

\vspace*{0.2in}

\ni FIG. 8.----~The binding energy of neutron stars as a function of stellar
gravitational mass for several EOSs listed in Table 1.  The
predictions of equation~(\ref{lybind}), due to Lattimer \& Yahil
(1989), are shown by the line labelled ``LY'' and the shaded region.

\vspace*{0.2in}

\ni FIG. 9.----~The binding energy per unit gravitational mass as a
function of compactness for the EOSs listed in Table 1.  Solid lines
labelled ``Inc'', ``Buch'' and ``T VII'' show predictions for an
incompressible fluid, the solution of Buchdahl (1967), and the Tolman
(1939) VII solution, respectively.  The shaded region shows the
prediction of equation~(\ref{newbind}).

\begin{figure}[hbt]
\epsfig{file=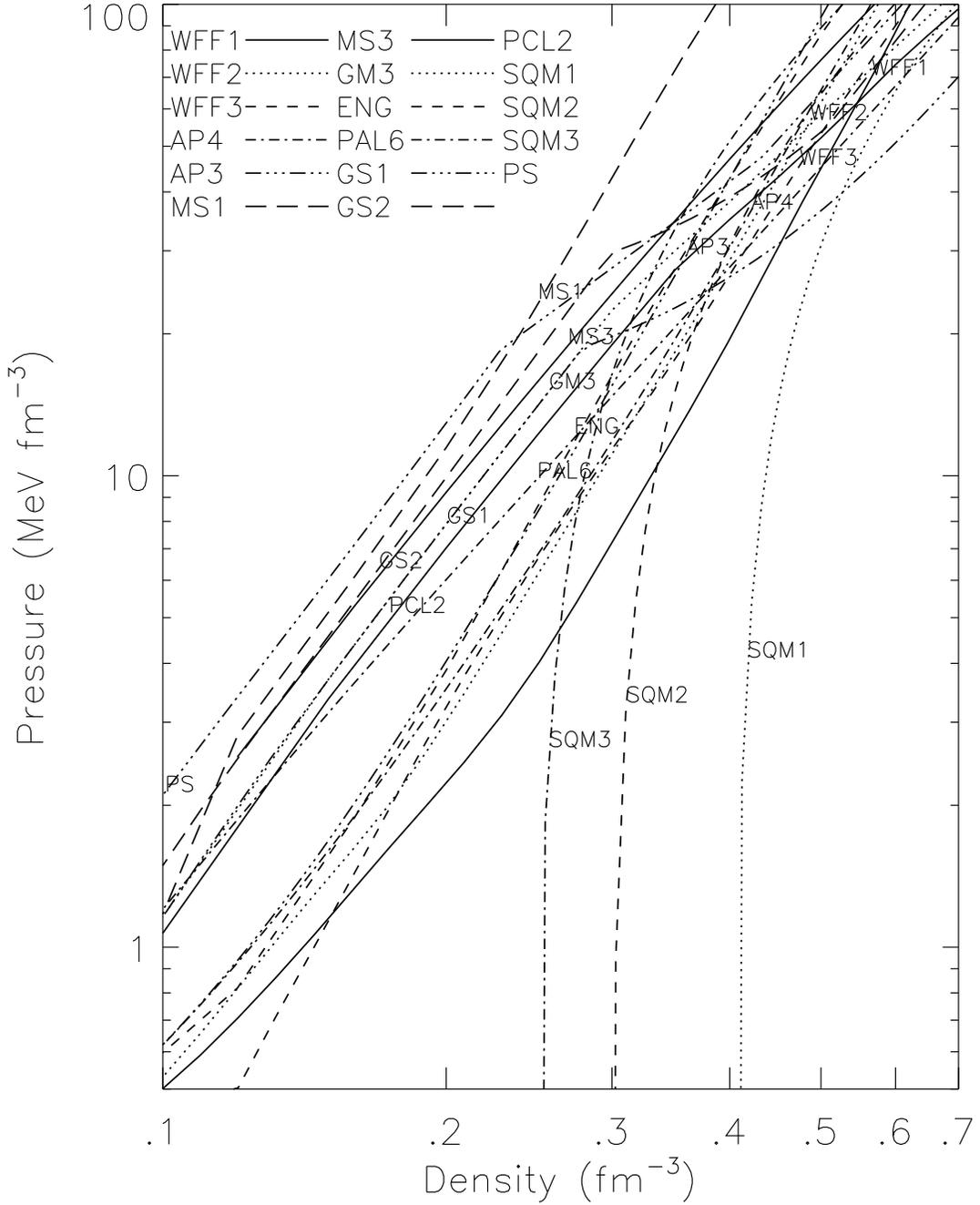, height=7.5in}
\caption{The pressure-density relation for a selected set of EOSs
contained in Table 1.  The pressure is in units of MeV fm$^{-3}$ and
the density is in units of baryons per cubic fermi.  The nuclear
saturation density is approximately $0.16$ fm$^{-3}$.}
\label{fig:P-rho}
\end{figure}

\begin{figure}[hbt]
\epsfig{file=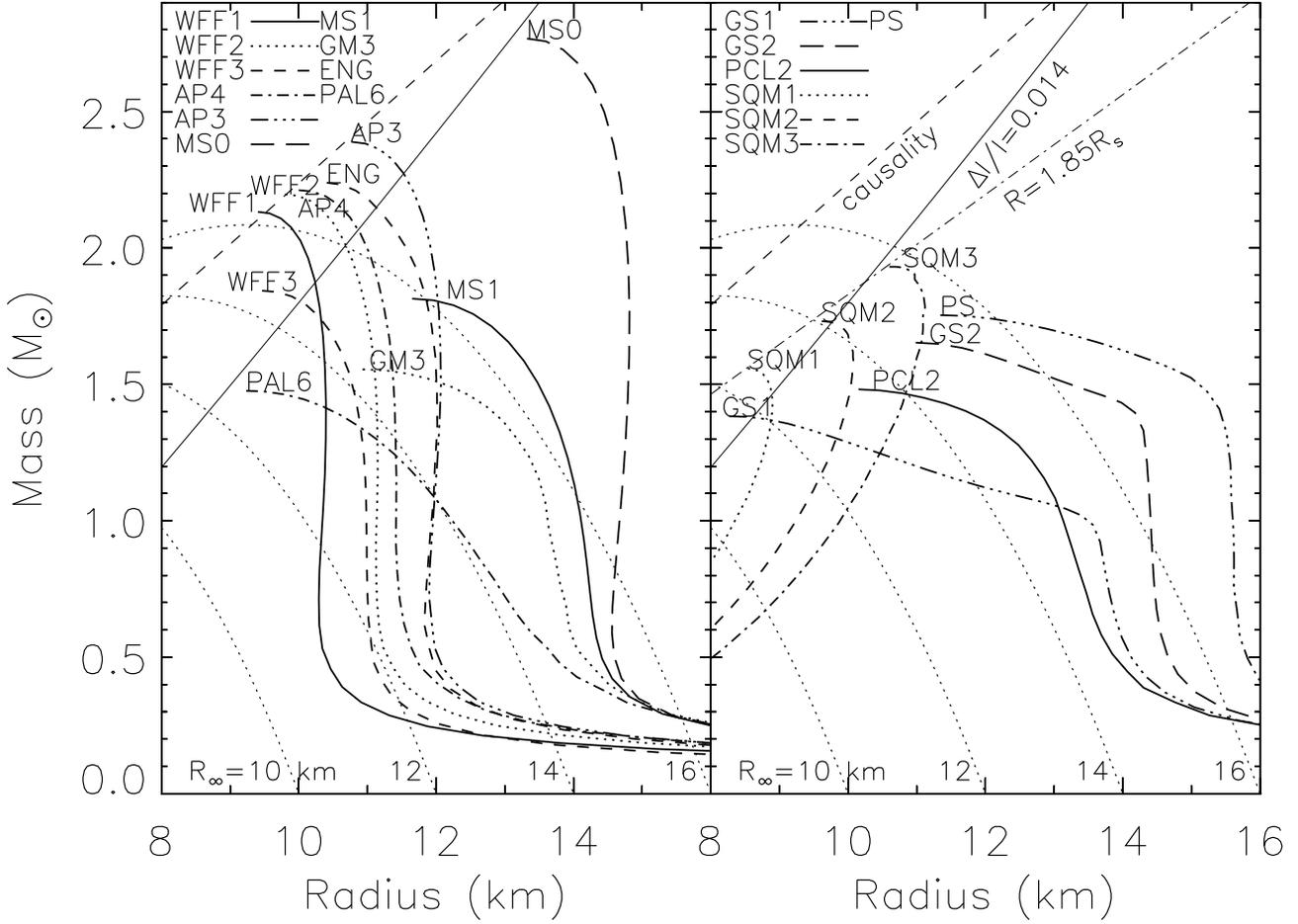, angle=90, height=5.5in}
\caption{Mass-radius curves for several EOSs listed in Table 1.  The
left panel is for stars containing nucleons and, in some cases,
hyperons.  The right panel is for stars containing more exotic
components, such as mixed phases with kaon condensates or strange
quark matter, or pure strange quark matter stars.  In both panels, the
lower limit causality places on $R$ is shown as a dashed line, a
constraint derived from glitches in the Vela pulsar is shown as the
solid line labelled $\Delta I/I=0.014$, and contours of constant
$R_\infty=R/\sqrt{1-2GM/Rc^2}$ are shown as dotted
curves. In the right panel, the theoretical trajectory of maximum
masses and radii for pure strange quark matter stars is marked by the
dot-dash curve labelled $R=1.85R_s$.}
\label{fig:M-R}
\end{figure}

\begin{figure}[hbt]
\epsfig{file=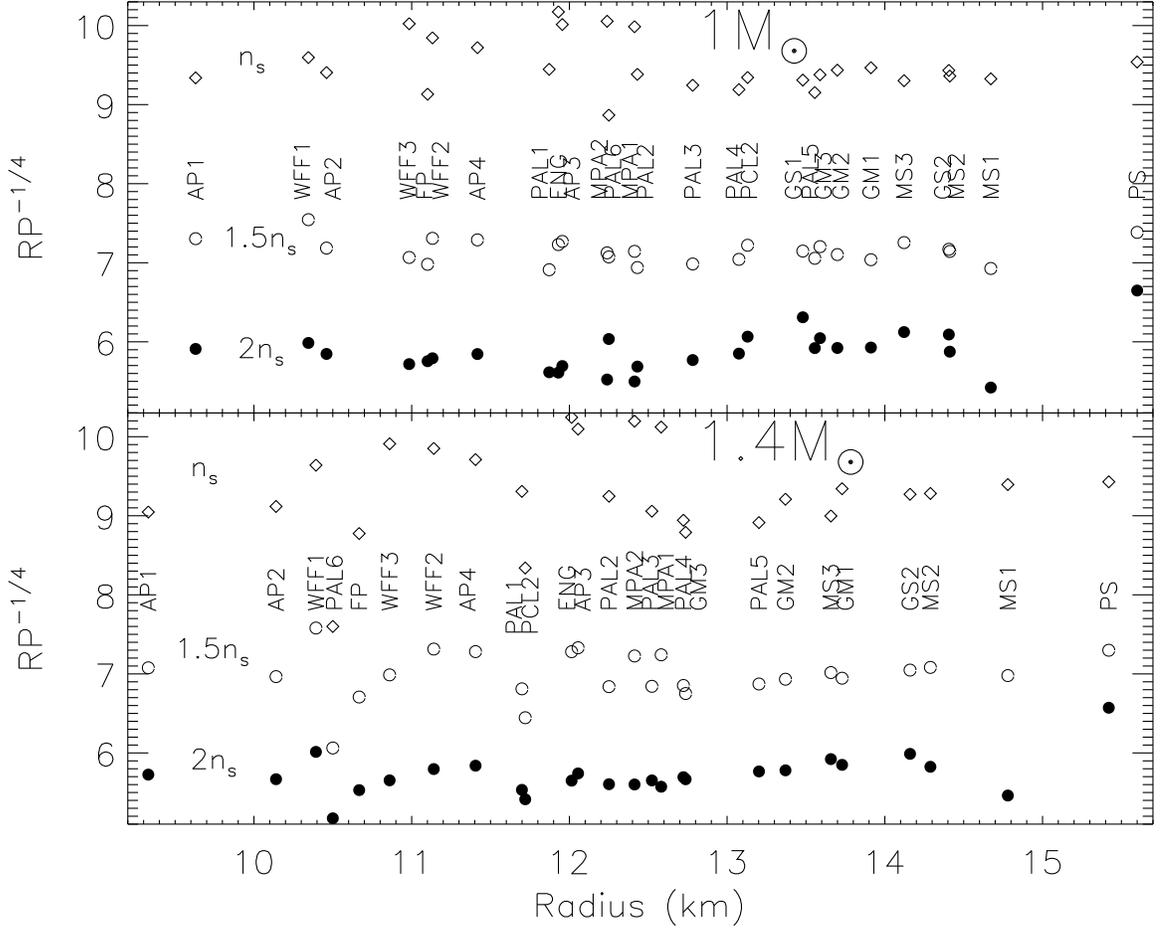, angle=90, height=5in}
\caption{Empirical relation between pressure, in units of MeV fm$^{-3}$,
and $R$, in km, for EOSs listed in Table 1.  The upper panel shows
results for 1 M$_\odot$ (gravitational mass) stars; the lower panel is
for 1.4 M$_\odot$ stars.  The different symbols show values of
$RP^{-1/4}$ evaluated at three fiducial densities.}
\label{fig:P-R}
\end{figure}

\begin{figure}[hbt]
\epsfig{file=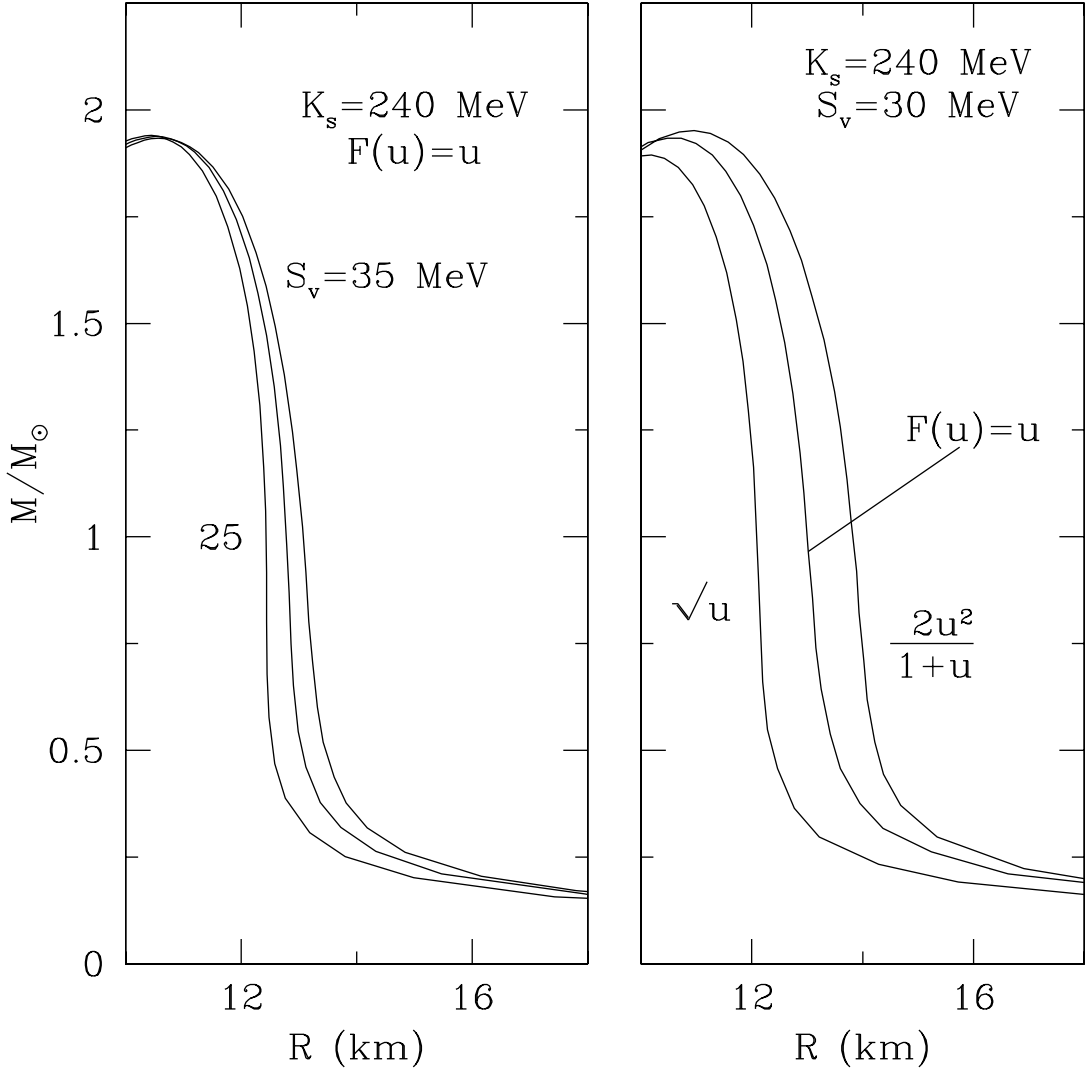, height=6in}
\caption{Left panel: Mass-radius curves for selected PAL (Prakash, Ainsworth
\& Lattimer 1988) forces showing the sensitivity to symmetry energy.
The left panel shows variations arising from different choices of
$S_v$, the symmetry energy evaluated at $n_s$; the right panel shows
variations arising from different choices of $S_v(n)$, the density
dependent symmetry energy.  In this figure, the shorthand $u=n/n_s$ is
used.}
\label{fig:alp9}
\end{figure}

\begin{figure}[hbt]
\epsfig{file=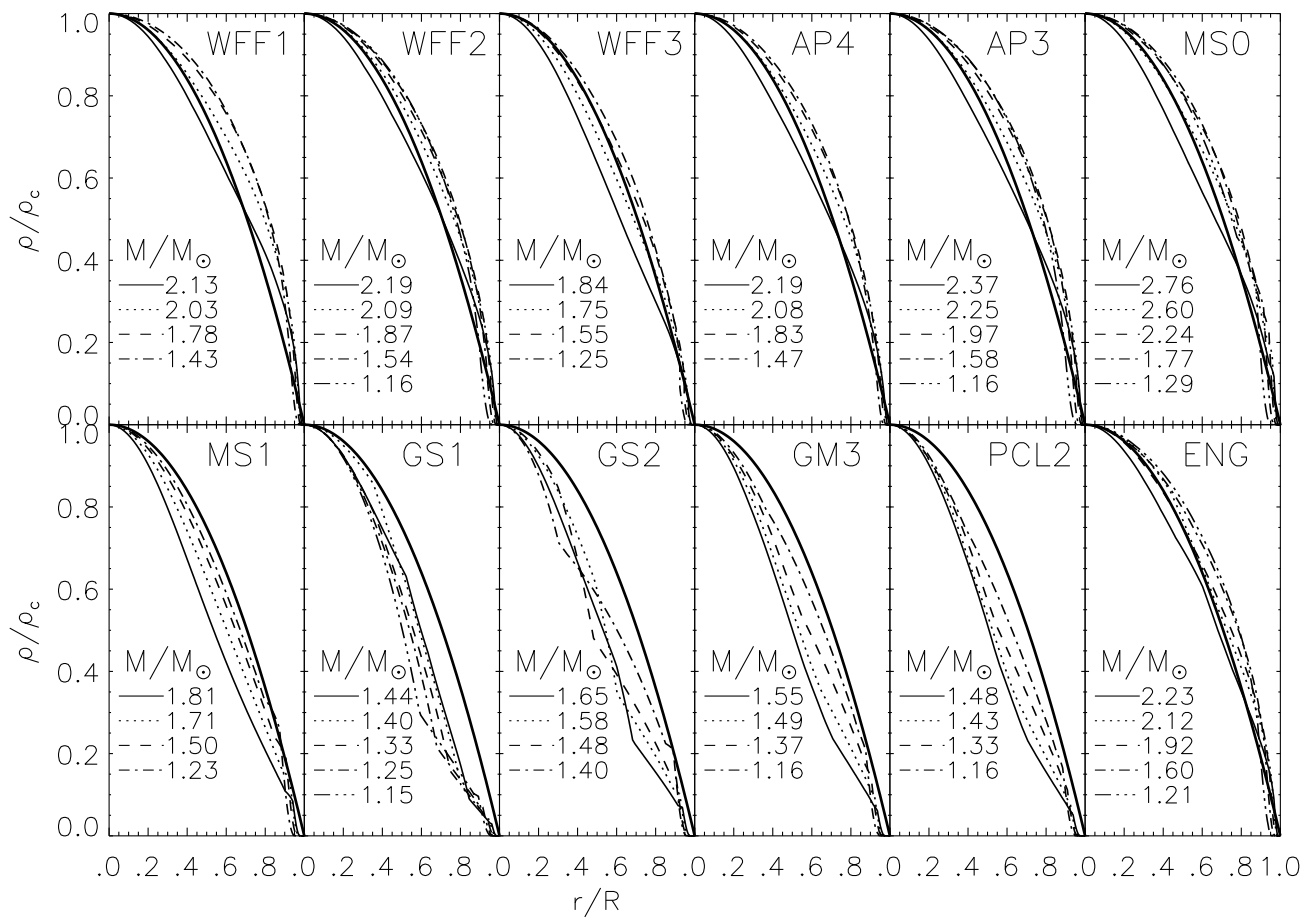, angle=90, height=5in}
\caption{Profiles of mass-energy density ($\rho$), relative to central
values ($\rho_c$), in neutron stars for several EOSs listed in Table 1.
For reference, the thick black lines show the simple quadratic
approximation $1-(r/R)^2$.}
\label{fig:prof}
\end{figure}

\begin{figure}[hbt]
\epsfig{file=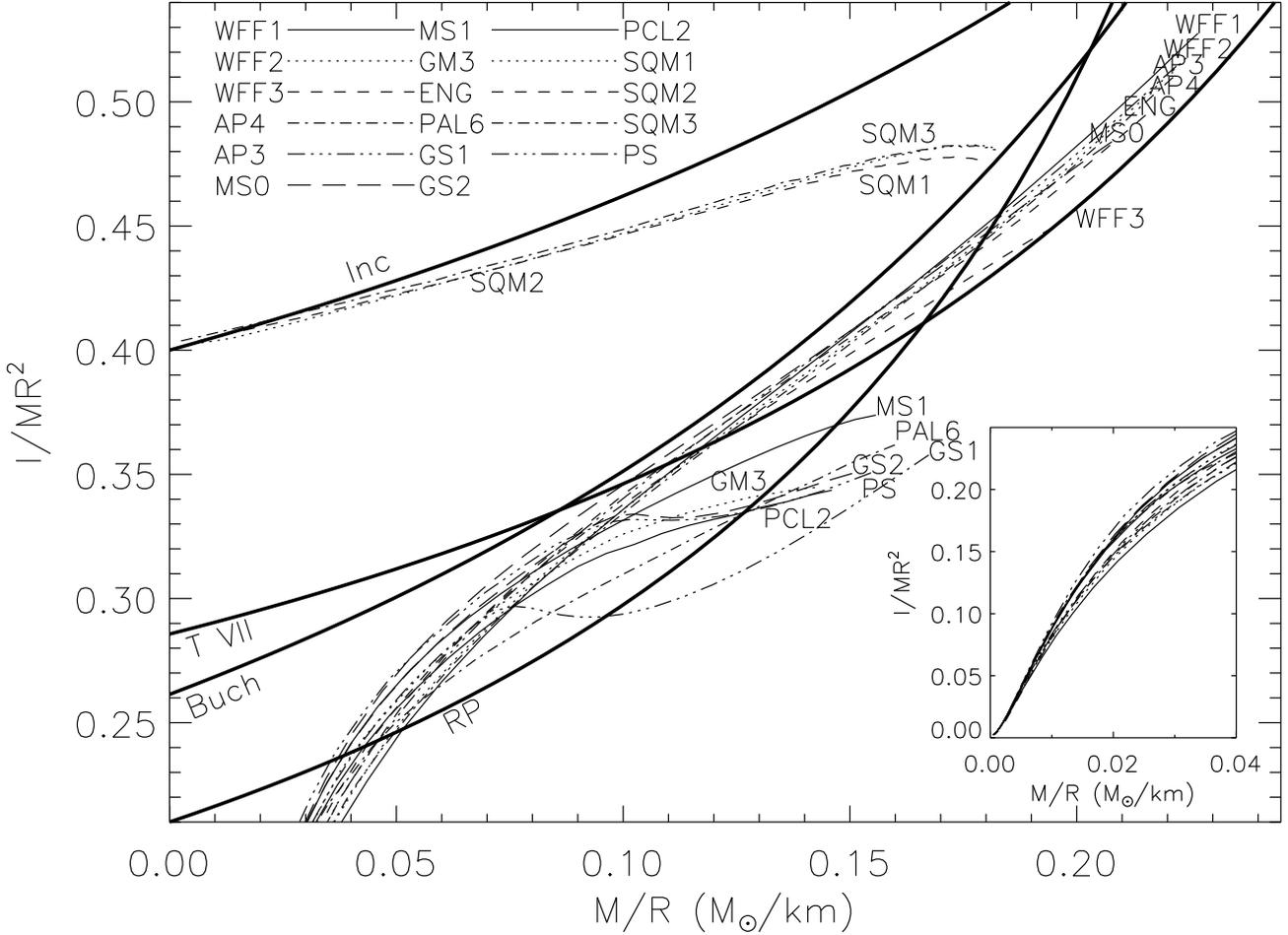, angle=90, height=5.5in}
\caption{The moment of inertia $I$, in units of $MR^2$, for several
EOSs listed in Table 1.  The curves labelled ``Inc'', ``T VII'', ``Buch''
and ``RP'' 
are for an incompressible fluid, the Tolman (1939) VII
solution, the Buchdahl (1967) solution, 
and an approximation of Ravenhall \& Pethick (1994), 
respectively.  The inset shows details of $I/MR^2$ for $M/R \rightarrow 0$.}
\label{mominert}
\end{figure}

\ni FIG. 10.---The pressures inferred from the empirical correlation
equation (\ref{correl}), for three hypothetical radius values (10, 12.5
and 15 km) overlaid on the pressure-density relations shown in Figure
~\ref{fig:P-rho}.  The light shaded region takes into account only the
uncertainty associated with $C(n,M)$; the dark shaded region also
includes a hypothetical uncertainty of 0.5 km in the radius
measurement.  The neutron star mass was assumed to be 1.4 M$_\odot$.

\begin{figure}[hbt]
\epsfig{file=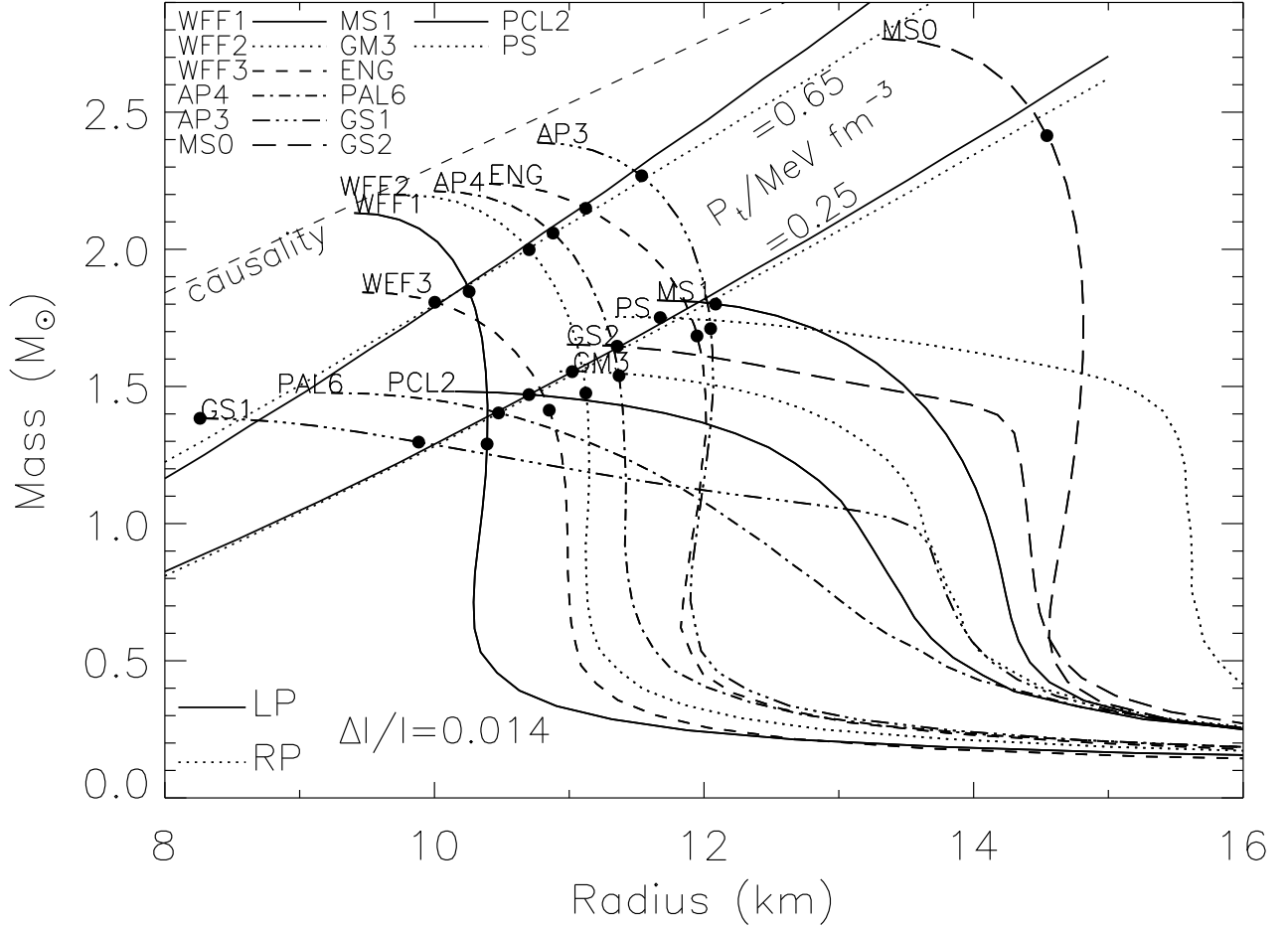, angle=90, height=5.5in}
\caption{Mass-radius curves for selected EOSs from Table 1, comparing
theoretical contours of $\Delta I/I=0.014$ from approximations
developed in this paper, labelled ``LP'', and from Ravenhall \&
Pethick (1994), labelled ``RP'', to numerical results (solid dots).
Two values of $P_t$, the transition pressure demarking the
crust's inner boundary, which bracket estimates in the literature, are
employed.  The region to the left of the $P_t=0.65$ MeV fm$^{-3}$
curve is forbidden if Vela glitches are due to angular momentum transfers
between the crust and core, as discussed in Link, Epstein \& Lattimer
(1999).  For comparison, the region excluded by causality alone lies
to the left of the dashed curve labelled ``causality'' as determined
by Lattimer et al. (1990) and Glendenning (1992).}
\label{fig:M-R-2}
\end{figure}


\begin{figure}[hbt]
\epsfig{file=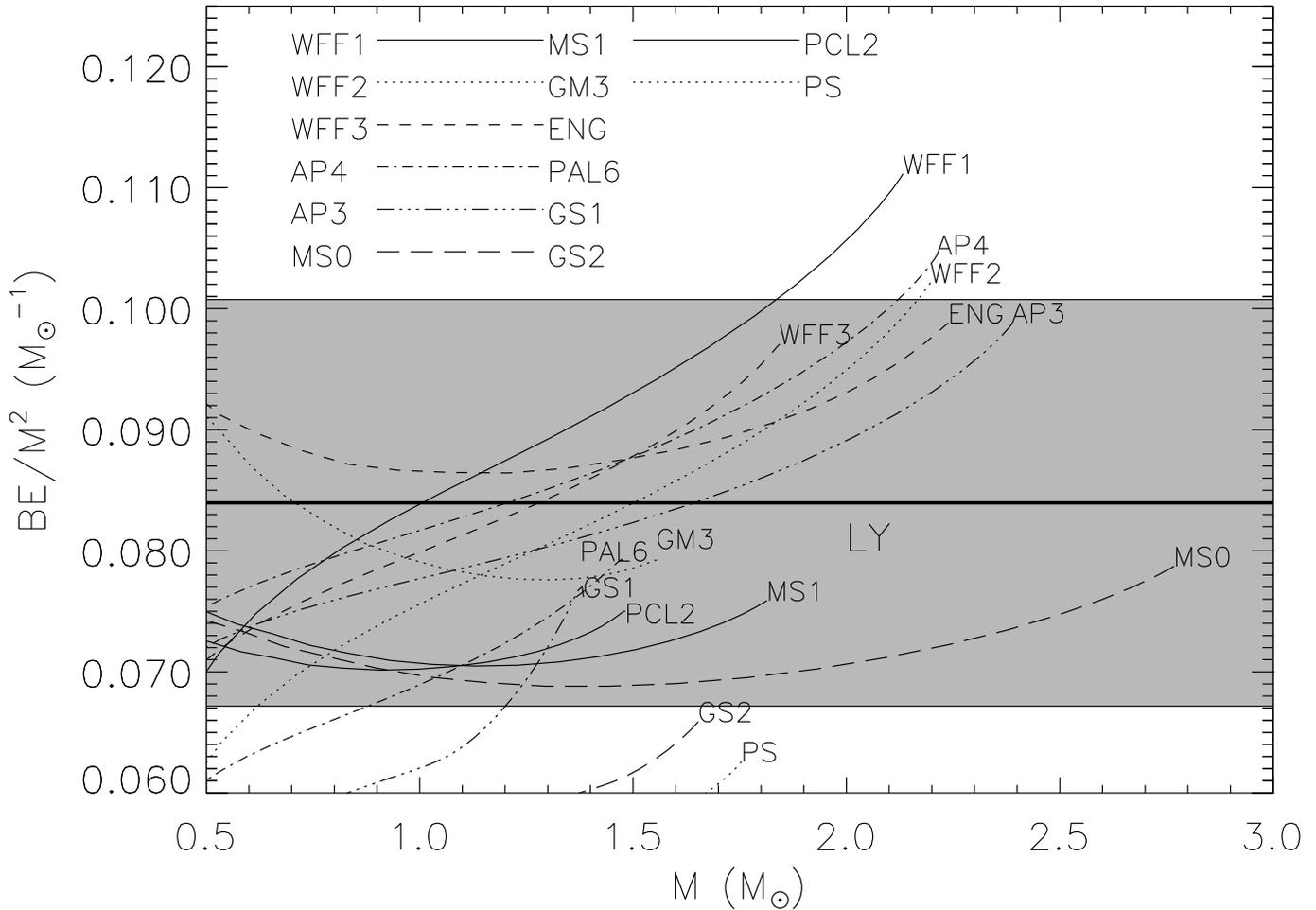, angle=90, height=5.5in}
\caption{The binding energy of neutron stars as a function of stellar
gravitational mass for several EOSs listed in Table 1.  The
predictions of equation~(\ref{lybind}), due to Lattimer \& Yahil
(1989), are shown by the line labelled ``LY'' and the shaded region.}
\label{bind}
\end{figure}


\begin{figure}[hbt]
\epsfig{file=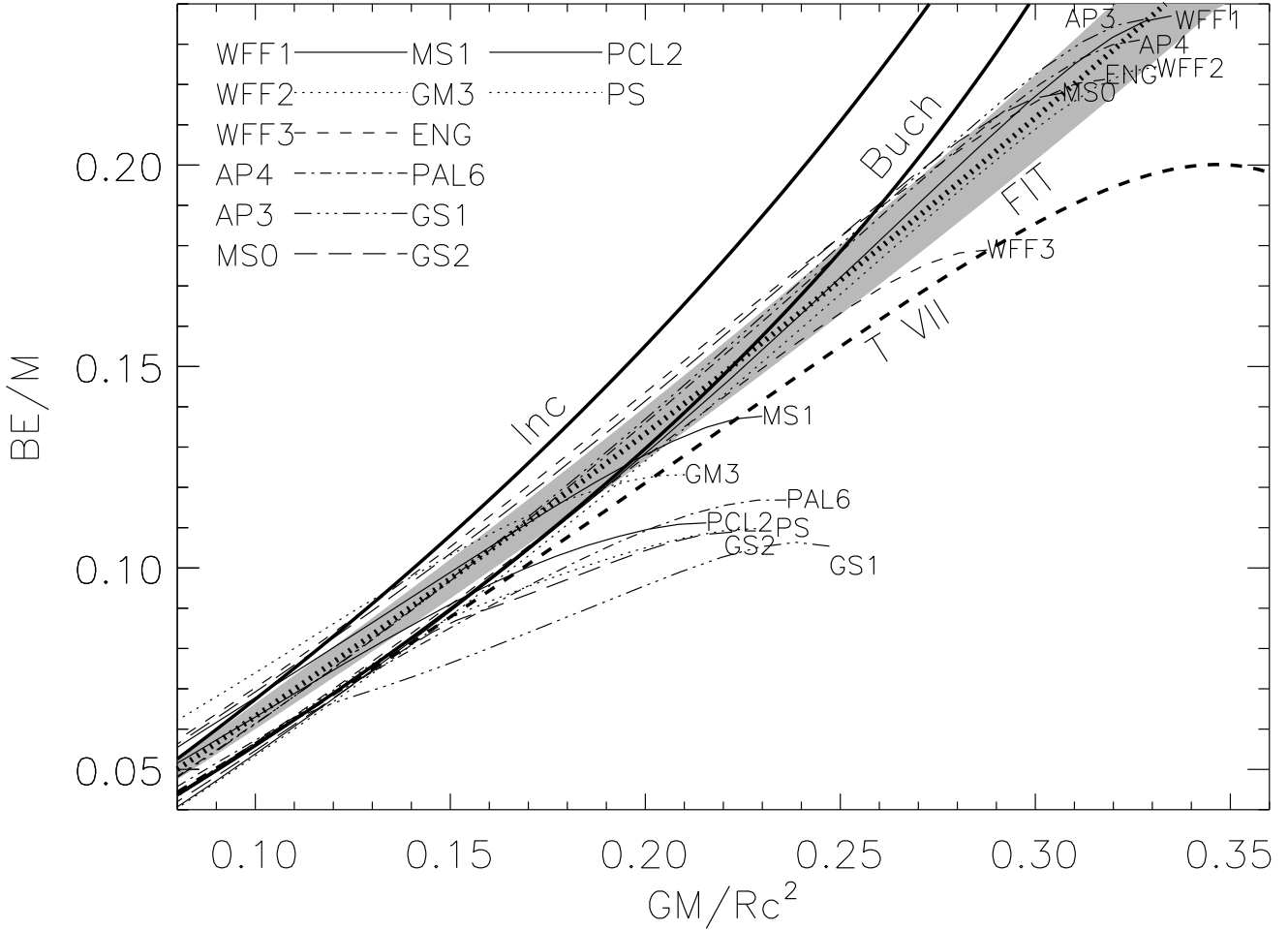, angle=90, height=5.5in}
\caption{The binding energy per unit gravitational mass as a function
of compactness ($\beta=GM/Rc^2$) for several EOSs listed in Table 1.
Solid lines labelled ``Inc'', ``Buch'' and ``T VII'' show predictions for an
incompressible fluid, the solution of Buchdahl (1967), and the Tolman (1939)
VII solution, respectively.  The dotted curve and shaded region
labelled ``FIT'' is the approximation given by equation~(\ref{newbind}).}
\label{bind1}
\end{figure}


\begin{figure}[hbt]
\epsfig{file=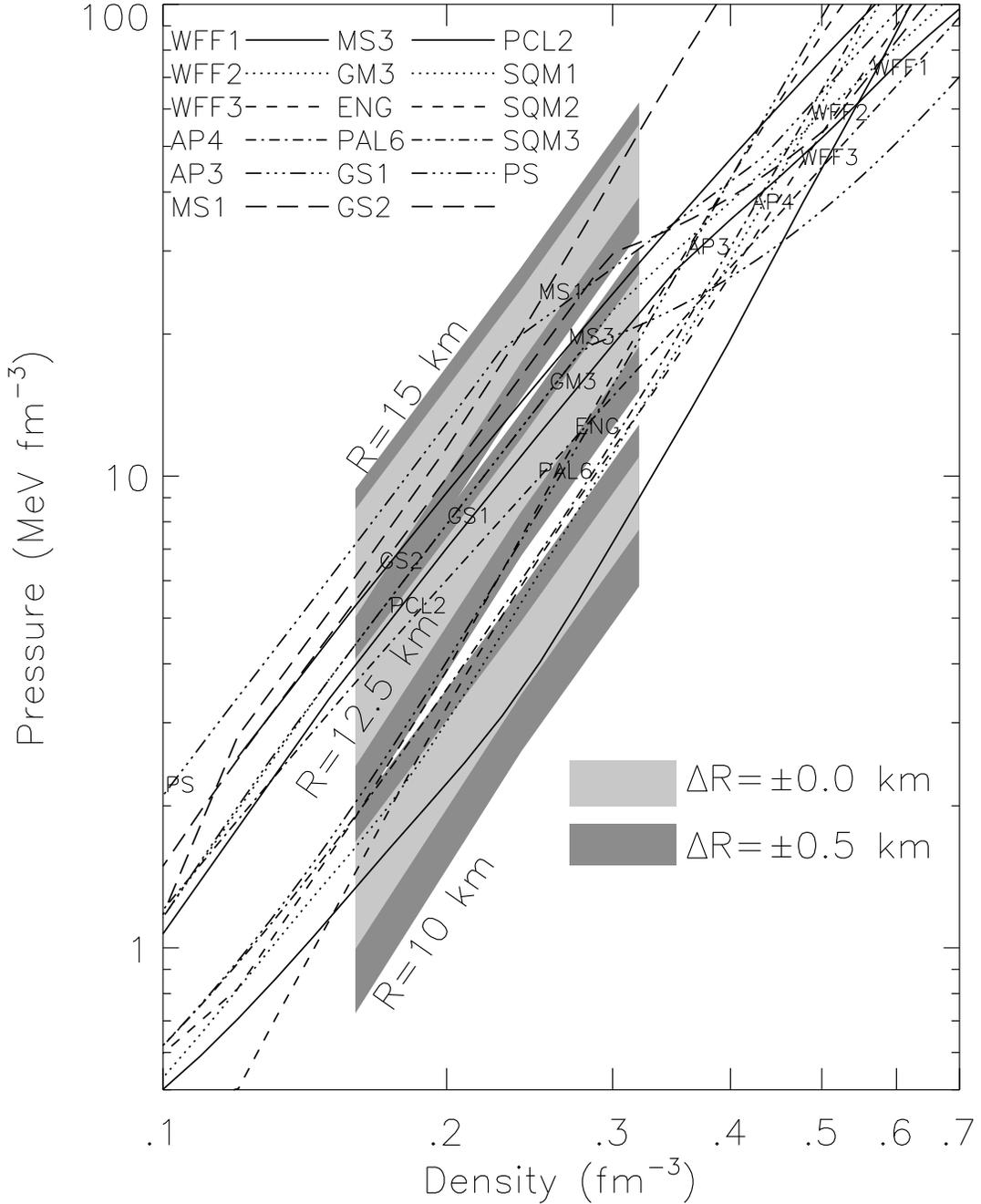, height=7.5in}
\caption{The pressures inferred from the empirical correlation
equation (\ref{correl}), for three hypothetical radius values (10, 12.5
and 15 km) overlaid on the pressure-density relations shown in Figure
~\ref{fig:P-rho}.  The light shaded region takes into account only the
uncertainty associated with $C(n,M)$; the dark shaded region also
includes a hypothetical uncertainty of 0.5 km in the radius
measurement.  The neutron star mass was assumed to be 1.4 M$_\odot$. }
\label{fig:err}
\end{figure}

\end{document}